\documentclass{jpp}

\usepackage{graphicx}
\usepackage{natbib}
\usepackage[colorlinks,urlcolor=blue,citecolor=blue,linkcolor=blue]{hyperref}
\usepackage{amsmath}

\ifCUPmtlplainloaded \else
  \checkfont{eurm10}
  \iffontfound
    \IfFileExists{upmath.sty}
      {\typeout{^^JFound AMS Euler Roman fonts on the system,
                   using the 'upmath' package.^^J}%
       \usepackage{upmath}}
      {\typeout{^^JFound AMS Euler Roman fonts on the system, but you
                   dont seem to have the}%
       \typeout{'upmath' package installed. JPP.cls can take advantage
                 of these fonts, if you use 'upmath' package.^^J}%
      }
  \else
  \fi
\fi

\ifCUPmtlplainloaded \else
  \checkfont{msam10}
  \iffontfound
    \IfFileExists{amssymb.sty}
      {\typeout{^^JFound AMS Symbol fonts on the system, using the
                'amssymb' package.^^J}%
       \usepackage{amssymb}%
       \let\le=\leqslant  
         \let\geq=\geqslant
      }{}
  \fi
\fi

\ifCUPmtlplainloaded \else
  \IfFileExists{amsbsy.sty}
    {\typeout{^^JFound the 'amsbsy' package on the system, using it.^^J}%
     \usepackage{amsbsy}}
    {}
\fi

\newcommand{\ce}{{\cal E}}

\newcommand{\oder}[2]{\frac{d #1}{d #2}}

\newcommand{\mtext}[1]{\quad\mbox{#1}\quad}

\newcommand{\be}{\begin{equation}} 
\newcommand{\ee}{\end{equation}}

\newcommand{\ba}{\begin{eqnarray}}
\newcommand{\ea}{\end{eqnarray}}

\newcommand{\Alfven}{Alfv\'{e}n }

\newcommand{\Bf}{{magnetic field}}

\newcommand{\Ef}{{electric  field}}
\newcommand{\Efs}{{electric fields}}

\newcommand{\EM}{electromagnetic}

\newcommand{\Lf}{Lorentz factor}

\renewcommand{\div}{{\rm \,div\,}}

\newcommand\etal{\textit{et al.\ }}
\newcommand\eg{\textit{e.g.,\ }}

\title[Explosive relativistic reconnection  and Crab Nebula flares]{Particle acceleration in explosive relativistic reconnection events and Crab Nebula gamma-ray flares}
\author[Maxim Lyutikov, Lorenzo Sironi,  Serguei Komissarov,  Oliver Porth]{Maxim Lyutikov$^1$, Lorenzo Sironi$^{2}$,  Serguei Komissarov$^{1,3}$,  Oliver Porth$^{3,4}$}
\affiliation{$^1$ Department of Physics, Purdue University, 
 525 Northwestern Avenue,
West Lafayette, IN
47907-2036, USA; lyutikov@purdue.edu
\\
$^2$  Department of Astronomy, Columbia University, 550 W 120th St, New York, NY 10027, USA; lsironi@astro.columbia.edu
\\
$^3$  School of Mathematics,
University of Leeds, LS29JT
Leeds, UK; s.s.komissarov@leeds.ac.uk
\\
$^4$  Institut f\"{u}r Theoretische Physik, 
J. W. Goethe-Universit\"{a}t, 
D-60438, Frankfurt am Main, Germany
porth@th.physik.uni-frankfurt.de
}

\begin{document}

\maketitle

\begin{abstract} 
We develop a  model of gamma-ray flares of the Crab Nebula  resulting from the magnetic reconnection events in highly-magnetized relativistic plasma. 
We first discuss physical parameters of the Crab nebula and review the theory of pulsar winds and  termination shocks.  We also review the principle points  of particle acceleration in explosive reconnection  events (Lyutikov \etal\ 2017a,b).

 It is required that particles producing flares are  accelerated in highly magnetized regions of the nebula. Flares  originate from the poleward regions at the base of Crab's polar outflow, where both the magnetization and the magnetic field strength are sufficiently  high.
The post-termination shock flow develops  macroscopic (not related to the plasma properties on the skin-depth scale) kink-type instabilities. The resulting  large-scales magnetic stresses  drive explosive reconnection events on the light-crossing time of the reconnection region. Flares are produced at  the initial stage of the current sheet development, during the X-point collapse. The model has all the ingredients needed for  Crab flares: natural formation of highly magnetized regions,   explosive dynamics on light travel time, development of high electric fields  on macroscopic scales  and  acceleration of particles  to energies well exceeding the average magnetic energy per particle.            
\end{abstract}

%\clearpage
%%%%%%%%%

%\tableofcontents

%sssssssssssssssssssssssssssssssssssssssssssss
%\section{Crab Nebula gamma-ray flares:  observational constraints and theoretical challenges}
\section{Introduction}
\label{intro}
%sssssssssssssssssssssssssssssssssssssssssssss

The Crab Nebula is an iconic astronomical object with an exceptionally long research history going back to the observations of 
a ``guest star'' by Chinese astronomers in 1054.  Astronomical studies of the nebula has proved to be an exceptionally fruitful both for  discovering new space phenomena and for developing of their physical models. As we now know the nebula has a lot in common with many other  cosmic objects, such as AGN jets, Gamma ray bursts, gamma-ray binaries etc and the knowledge gained from its studies has found numerous applications in many other arias of astrophysics.    

It is now well established that the nebula is a result of the interaction between the magnetized ultra-relativistic wind from the Crab pulsar and the supernova shell  ejected during the 1054 explosion. The physical models of this interaction has evolved from simplified spherically-symmetric 1D analytic models \citep{reesgunn,kc84}  through early axisymmetric  numerical models \citep{KomissarovLyubarsky,DelZanna04,2016MNRAS.456..286L}   to fully 3D sophisticated numerical simulations  \citep{2013MNRAS.431L..48P,2016JPlPh..82f6301O}.  The current fluid models seem to explain many of both the large-scale and fine features of the nebula quite convincingly but there are still many uncertainties. The most important of them concerns the origin of the high-energy leptons producing the observed non-thermal emission of the nebula.         

Following the pioneering work by \citet{reesgunn} and \citet{1984ApJ...283..710K}, it is usually assumed that these particle are accelerated at the termination shock of the pulsar wind. This assumption seems rather natural given the success of the particle acceleration theory for non-relativistic shocks \citep[\eg][]{2011ApJ...733...63R}. However, in the case of pulsar winds we are dealing with relativistic shocks, which are much less promising as particle accelerators in the presence of large-scale magnetic field.  According to the recent PIC simulations \citep{SironiSpitkovsky09} this acceleration mechanism operates only for very low magnetization of upstream plasma \citep[$\sigma<10^{-3}$][]{2005AIPC..801..345S,2009ApJ...698.1523S}. 

Unfortunately, the magnetization of pulsar winds cannot be measured directly and we have to deal with indirect estimates and theoretical predictions only. These do not always go hand-in-hand. Thus the theory of pulsar magnetospheres and ideal MHD winds predict very high magnetization of pulsar winds whereas the original spherically symmetric model of the  Crab nebula fits the data only if the magnetization is very low. This controversy is known as the $\sigma$-problem.   Interestingly, the magnetization deduced by \citet{kc84} marginally satisfies the above condition of the shock acceleration mechanism.  

The brightest and the most compact resolved feature of the Crab nebula is the so-called inner knot located with one arc-second from the Crab pulsar. A similar knot was found in the synthetic images of the Crab nebula based on the low-$\sigma$ wind models  \citep{KomissarovLyubarsky}, where it was identified with Doppler-boosted emission of the termination shock.    \cite{2015MNRAS.454.2754Y} and \cite{2016MNRAS.456..286L} explored the shock model of the knot in details and concluded that it is consistent with the observations only if the downstream plasma magnetization is low.  For an ideal fast-magnetosonic shock, this also means low upstream magnetization.  If however the upstream magnetic field is highly inhomogeneous things can be rather different. In this case the shock can trigger rapid magnetic dissipation immediately downstream of the ideal shock and the ``duet'' of  this shock and the post-shock dissipation layer emulates an ideal shock in weakly-magnetized plasma \citep{2001ApJ...547..437L,2011ApJ...741...39S}. The dissipation layer could be the site of not only plasma heating but also acceleration of non-thermal particles. However, the recent PIC simulations of termination shocks find that broad non-thermal tails of are produced (via standard Fermi mechanism) only in a very narrow equatorial section of the termination shock where the wind-averaged magnetic field is below about one hundredth of its typical stripe value. Elsewhere, a relativistic Maxwellian spectrum is expected instead \citep{2011ApJ...741...39S}. This may explain why inner knot is not visible both in the radio and X-ray bands \citep{2013ApJ...765...56W,2015ApJ...811...24R}.          

The magnetic dissipation in the striped zone cannot completely solve the $\sigma$-problem unless the zone fills the whole wind \citep[the magnetic inclination angle of the pulsar $\alpha\approx\pi/2$;][]{Coroniti90,2013MNRAS.428.2459K}. If however  $\alpha<\pi/2$ then inside 
the polar cones of the opening angle $\theta_p\approx \pi/2-\alpha$, the wind's magnetic field is unidirectional and hence it is not dissipated at the termination shock as in the striped wind zone. Instead it is injected into the nebula with a highly relativistic speed where it has to dissipate in some other way in order to accommodate the observed mean value of the magnetic field in the Crab nebula \citep{2013MNRAS.428.2459K}.  \cite{2013MNRAS.431L..48P} explored this idea using 3D RMHD simulations. Their wind model accommodated the existence of both the unidirectional polar and the striped equatorial zones but due to the scheme limitations they could only consider winds with mean magnetization up to $\sigma\approx 1$. This value is three orders of magnitude above the one deduced in the 1D model of  \citet{kc84} but this had a little impact on the basic parameters of the produced nebula, all because of the strong magnetic dissipation in its body. Thus magnetic dissipation inside nebula appears to be the key to the solution of the $\sigma$-problem. 
%The results of recent 1D modelling with ad hoc magnetic dissipation strengthen this conclusion (Shuta J. Tanaka, talk at ``Physics of Neutron Stars'', St.Petersburg, 2017).   

The impact of magnetic dissipation on the mean spectrum of the Crab nebula remains to be identified.  At present, the most popular model of the nebula radiation is based on the shock model by  \citet{kc84}.  As we have just discussed, it may be completely wrong but it describes rather well the broad-band synchrotron spectrum of the nebula (except the radio emission).  However the observed cutoff of the spectrum at  $\sim 100$ MeV is  very close to the upper limit imposed by the balance of lepton acceleration and synchrotron  losses (see below). \cite{2010MNRAS.405.1809L} argued that this is inconsistent with any  stochastic acceleration mechanism, including the shock acceleration, the conclusion confirmed by the self-consistent PIC simulations of unmagnetized shocks \citep{2013ApJ...771...54S}.  This is another factor that undermines the Kennel-Coroniti model of the Crab radiation. 

The recent discovery of gamma ray flares \citep{2011Sci...331..736T,2011Sci...331..739A,2012ApJ...749...26B} adds another twist to the story of particle acceleration in the nebula. These flares last for a very short time, from few days to a couple of weeks. Moreover, the flux is seen to vary on the timescale down to few hours.  
When flaring, the gamma-ray flux may increase by up to two orders of magnitude above the normal level. For the most powerful event to date, the April 2011 flare, the isotropic luminosity reached  $L_{max}= 4\times 10^{36}$erg/s at its peak.    The flux variations are detectable in a rather short range of photon energy, from 70~Mev to 2~GeV, which is at the junction of the synchrotron and inverse-Compton ``humps'' of the nebula spectrum. The flaring component can be modeled as a power law with an exponential cut-off at the high-energy end 
\be
     dn_\nu/d\ce  \propto \ce^{-a} \exp(-(\ce/\ce_c)^\kappa)
     \label{flare_sp}
\ee
with $a=1.27\pm 0.12$ and the cut-off energy reaching $\ce_{c,max}\approx 500\,$MeV at the peak of the April 2011 flare \citep{2012ApJ...749...26B}. Provided the radiation is produced via the synchrotron mechanism, which is not in doubt, the exceptionally high value of $\ce_{c,max}$ presents a clear challenge to any particle acceleration mechanism \cite{2010MNRAS.405.1809L}. The resolution of the gamma-ray telescopes is too low to pinpoint the location of the flares in the nebula. In spite of substantial efforts to identify the flares with events in other energy bands, where the resolution is much higher, no progress has been made so far in this direction \citep{2013ApJ...765...56W}.

If we leave aside the difficulties of particle acceleration at relativistic shocks and stick to the Kennel-Coroniti model, then given the short synchrotron cooling time we would expect the gamma-ray emission to be generated only in the very vicinity of the termination shock where the flow is still highly relativistic and hence we would expect it to be strongly Doppler-beamed.  In fact, the observed synchrotron gamma-ray would be dominated by the contribution of the inner knot \citep{2011MNRAS.414.2017K} and this would be a natural cite for the gamma-ray flares.  Indeed, its linear size is only few light days, which is comparable to the duration of flares, and its emission is blue-shifted, which helps to alleviate the problem with the peak photon energy. 
The variability could be driven by the violent dynamics of the inner nebula resulting in rapid changes of the shock geometry \citep{2011MNRAS.414.2017K,2012MNRAS.422.3118L}. However, such a mechanism predicts a correlated variability in all energy bands  which is not observed \citep{2015ApJ...811...24R}.

Alternatively, the flares could be associated with magnetic reconnection events in the highly magnetized 
relativistic plasma of the nebula \citep{2010MNRAS.405.1809L,2014PhPl...21e6501C,2013ApJ...770..147C,2016ApJ...828...92Y,2017ApJ...847...57Z}. 
The particle acceleration during relativistic magnetic reconnection in pair plasma has been addressed in a number of studies, both in general \citep[\eg][and others]{2012ApJ...750..129B,2014ApJ...783L..21S,2015ApJ...806..167G} and with a particular focus on the Crab nebula \citep[][ papers I and II]{2012ApJ...746..148C,2012ApJ...754L..33C,2014PhPl...21e6501C,2013ApJ...770..147C}. 
They have demonstrated that the relativistic magnetic reconnection is a promising model for the Crab flares. In particular, 1) the emerging particle spectrum is a hard power law with the spectral index $a_e<2.0$ for $\sigma>10$; 2) the top energy of photons can somewhat exceed the classical 
radiation reaction limit of $150\,$MeV without employing the Doppler beaming and 3) the accelerated particles form wandering narrow beams which helps to explain the observed short time-scales of flare variability.      

The typical initial configuration adopted in many reconnection studies, including those directly addressing the nature of the gamma-gay flares, is a plane Harris current layers of microscopic thickness (e.g. the electron skin depth scale). Such a setup does not allow to tackle a number of important questions concerning the flares. How do such current sheets form in the first place?  What are the explosive dynamic processes which result in the flares? What determines the duration of these flares?  Here we attempt to combine the conclusions made in these studies with the those accumulated over the years of studying pulsars and their nebulae into the theory of Crab's flares.  We also discuss their implications for the theory magnetic reconnection in highly magnetized plasma in general.

\newcommand{\fracp}[2]{\left(\frac{#1}{#2}\right)}

%sssssssssssssssssssssssssssssssssssssssssssss
\section{Crab pulsar, its wind and nebula}
\label{PN}
%sssssssssssssssssssssssssssssssssssssssssssss

In order to evaluate the potential of the magnetic reconnection model of the Crab flares and figure out the possible location of these events we start by reviewing the parameters of the Crab's pulsar wind and its nebula. 

%ssssssssssssssssssss
\subsection{The wind}
\label{wind}

In the wind, the magnetic field is dominated by the azimuthal component, whose magnitude varies as 
\be
    B_\phi (r,\theta) = B_{lc} \fracp{r_{lc}}{r} \sin^n\theta\,, 
    \label{B-wind}
\ee   
where $r_{lc} =c/\Omega$ is the radius of light cylinder, $B_{lc}$ is the typical value of $B$ at the light cylinder and $\Omega$ is the angular velocity of the pulsar. In the monopole model of the pulsar magnetic field $n=1$ but there is no easy way of finding its value in the more realistic the dipolar model.  Recent numerical simulations suggest that it could be $n=2$ 
\citep{2016MNRAS.457.3384T}.  This magnetic field is unidirectional in the polar region $\theta < \theta_p$ but in the equatorial region it comes in the form of stripes with opposite magnetic directions. In the monopole model, $\theta_p=\pi/2-\alpha$, where $\alpha$ is the angle between the magnetic and rotational axes of the pulsar but in the dipolar model the poloidal field lines are not straight and this relation may no longer hold. 

\begin{figure}[htbp]
\begin{center}
\includegraphics[width=0.4\textwidth]{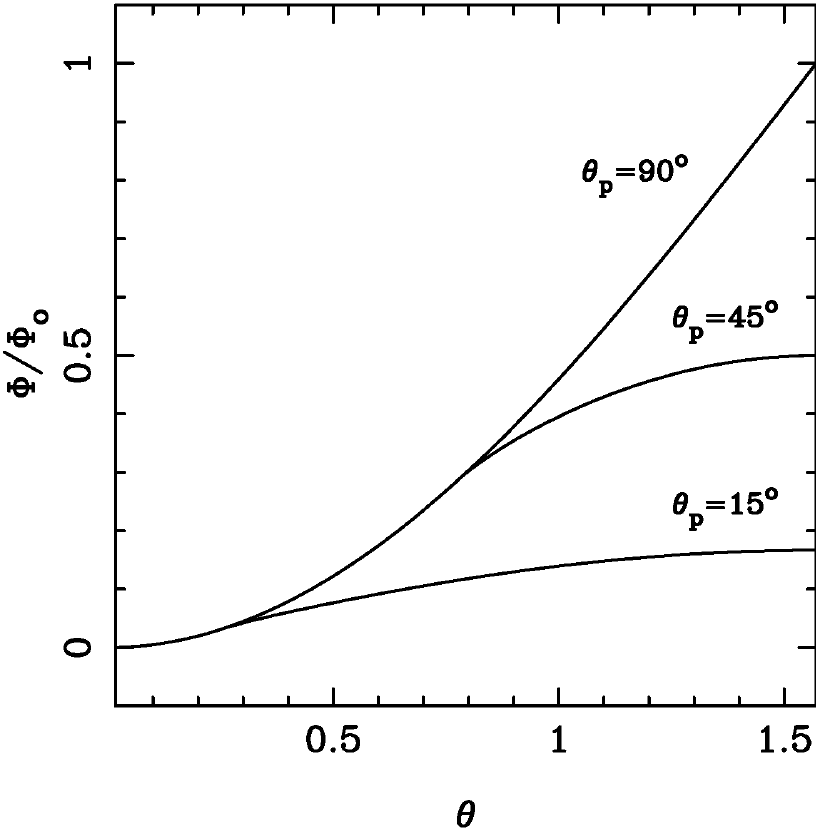}
\includegraphics[width=0.4\textwidth]{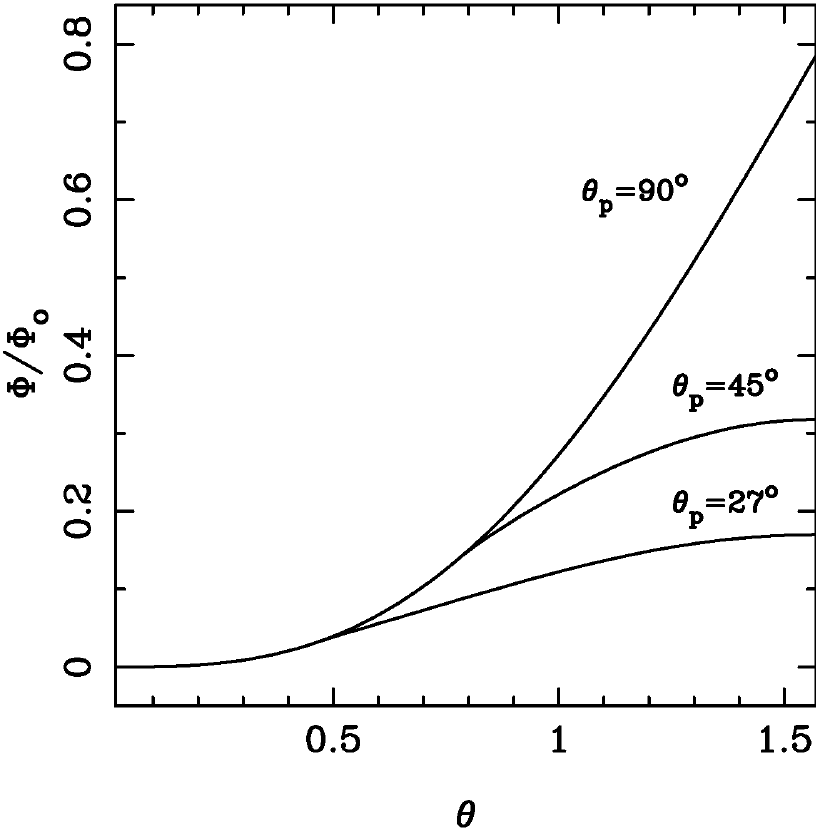}
\caption{Available potential across the winds depending on the size of the polar zone. Left panel shows the results for $n=1$ and the right panel for $n=2$}
\label{fig:potential}
\end{center}
\end{figure}

For $r\gg r_{lc}$ the wind becomes relativistic and radial and its electric field is dominated by the polar component 
\be 
     E_\theta = (v/c) B_\phi = B_\phi \,.
\ee     
The corresponding potential drop across the flow in the polar zone is
\be 
   \Phi(\theta) = \int\limits_0^\theta E_\theta r d\theta = \Phi_0 f(\theta) \,,
\label{pot-drop}
\ee 
where
\be
    \Phi_0=B_{lc} r_{lc}
\ee
and 
\be
   f(\theta)= \left\{ 
   \begin{array}{ll}
        1-\cos(\theta) & \mbox{for } n=1 \\
        0.5(\theta - 0.5\sin2\theta) & \mbox{for } n=2 
   \end{array}
    \right. .
\label{var-n}
\ee
In the striped wind zone one has to take into account the alternation of magnetic field direction and use the stripe-averaged magnetic field, which is smaller than that in Eq.(\ref{B-wind}) (see \citet{2013MNRAS.428.2459K}). As a result, inside the striped wind zone the potential grows much slower (see Figure~\ref{fig:potential}).  

The corresponding Poynting vector in the wind
\be
        S_P = \frac{1}{4\pi} E_\theta B_\phi c = \frac{c}{4\pi} \fracp{\Phi_0}{r}^2 \sin^{2n}\theta \, 
\ee
which leads to the total wind power 
\be 
     L_w = a_n c\, \Phi_0^2 \, 
\ee
where $a_1=2/3$ and $a_2=8/15$\,. Thus one can find $\Phi_0$ given the measurement of $L_w$ as the spindown power of the pulsar. For the Crab pulsar $L_w \approx 4\times10^{38}\mbox{erg}\,\mbox{s}^{-1}$ which leads to 
$\Phi_0 \approx 42\,\mbox{PeV}$, the corresponding electron Lorentz factor  
\be
    \gamma_{max} \approx 8.2\times 10^{10} 
    \label{gammamax1}
\ee
and the wind magnetic field 
\be
   B_\phi \approx 5\times10^{-2} r_{,ld}^{-1} \sin^n\theta\, \mbox{G} \,,
   \label{B_phi}
\ee
where $r_{,ld}$ is the distance measured in light days. 

At the light cylinder, the Goldreich-Julian number density of charged particles 
\be 
    n_{GJ} = \frac{1}{4\pi e} \div{E} \approx \frac{1}{4\pi e} \frac{B_{lc}}{r_{lc}} = \frac{1}{4\pi e} \frac{\Phi_0}{r_{lc}^2} 
\ee
whereas the actual density governed by the pair-production rate $n=\lambda n_{GJ}$, where $\lambda$ is called the multiplicity factor. The corresponding total $e_{\pm}$-particle flux of the wind 
\be
     \dot{N} \approx 4\pi r_{lc}^2 n c = \frac{\Phi_0 c}{e} \approx 8.7\times10^{37} \lambda_4\, \mbox{s}^{-1} \,, 
     \label{Ndot-theory}
\ee   
where $\lambda_4 = 10^{-4}\lambda$. This corresponds to the particle density in the wind
\be
     n \approx 3.4\times10^{-5} \lambda_4 r_{,ld}^{-2}\,\mbox{cm}^{-3} \,.
\label{wind-n}
\ee

Along the particle flux tube of a steady-state wind, the particle flux $S_\pm$ and the total energy flux $S_{tot}=S_P+m_ec^2 \gamma_w h S_\pm$, where $\gamma_w$ is the wind Lorentz factor and $h=w/\tilde{n}m_e c^2$ is the relativistic enthalpy per unit particle rest mass,  are conserved. This  leads to the integration constant along streamlines 
\be 
     \sigma_M = \frac{S_{tot}}{S_\pm m_e c^2} =  \gamma_w h (1+\sigma)\,,
\ee 
where $\sigma=S_P/S_p$ is the ratio of the Poynting flux and the particle energy flux $S_p=S_\pm  m_e c^2 h\gamma_w$. 
In the theory of pulsar winds, $\sigma_M$ is known as Michel's sigma \citep{1969ApJ...158..727M} but in the theory of   
magnetically-dominated jets  it is simply denoted as $\mu$ \citep[e.g.][]{2007MNRAS.380...51K}.
Replacing $S_P$ and $S_\pm$ with their densities, $B_\phi^2 c/4\pi$ and $nc$ respectively, we obtain 
\be
    \sigma= \frac{B_\phi^2}{4\pi m_e c^2 h n \gamma_w} = \frac{\tilde{B}_\phi^2}{4\pi m_e c^2 h \tilde{n}} \,,
\ee
where $\tilde{B}$ and $\tilde{n}$ are measured in the wind frame. Using the parameters at the light cylinder to estimate $\sigma_M$, we find 
\be 
   \sigma_M = \frac{\Phi_0 e}{m_e c^2 \lambda} \approx 8.2\times10^6 \lambda_4^{-1}\,.
   \label{sigma-m}
\ee  
For a cold wind $h=1$ and hence
\be 
    \gamma_w = \frac{\sigma_M}{1+\sigma} \,. 
    \label{gam-sig}
\ee
Thus the total conversion of the Poynting flux into the kinetic energy of the bulk motion yields the wind speed   
$$
    \gamma_{w}^{max} = \sigma_M \,.
$$   
However in the limit of ideal MHD, reaching this  asymptotic value is quite problematic.  Indeed, the wind must first become super-fast-magnetosonic.  At the transonic point  
$$
      \gamma_w=\sqrt{\sigma} \,
$$
\citep[e.g][]{2012MNRAS.422..326K}. Combining this result with Eq. (\ref{gam-sig}) we find that at the fast magnetosonic point 
\be
     \gamma_{w,f} = \sigma_M^{1/3} = 200\,\lambda_4^{-1/3}  \quad\mbox{and}\quad \sigma_f = 4\times10^4\lambda_4^{-2/3} \,.
     \label{sigmaf}
\ee
Beyond this point the wind becomes causally disconnected, which strongly reduces the efficiency of the ideal collimation-acceleration mechanism  \citep[e.g. ][]{1998MNRAS.299..341B,2009MNRAS.394.1182K}.    Thus we do not expect the wind's Lorentz factor to exceed the value given in Eq.(\ref{sigmaw}) by more than a factor of ten. Using the fast-magnetosonic Mach number  $M\approx \Gamma_w/\sqrt{\sigma}$, the wind parameters at the termination shock can be estimated as 
\be
     \gamma_{w,u} = M^{2/3} \sigma_M^{1/3} = 200\,M^{2/3}\lambda_4^{-1/3}  
     \quad\mbox{and}\quad \sigma_u = (\sigma_M/M)^{2/3} = 4\times10^4 M^{-2/3} \lambda_4^{-2/3} \,. 
     \label{sigmaw}
\ee
Strictly-speaking, these results do not apply to the striped wind zone where the magnetic dissipation of its stripes may come into play and allow some additional acceleration of the flow.  Even in the polar zone things can be more complicated if a substantial fraction of the energy is carried away by fast magnetosonic waves emitted into the zone because of the misalignment of the magnetic and rotational axes of the Crab pulsar. \citet{2003MNRAS.339..765L} argues that for $\lambda\gtrsim 10^7$ these waves steepen into shocks and dissipate before reaching the termination shock and the generated heat is quickly converted into the bulk kinetic energy of the wind particles. This process would set an upper limit on the wind magnetisation $\sigma$ at the termination shock equal to the inverse fraction of the energy flux carried by the waves and the Poynting flux of the wind.  For example, if the waves contribute only one percent to the wind energy budget then $\sigma_u<100$. For smaller multiplicity,  the waves would dissipate only after crossing the termination shock and the dissipation rate increases sharply in the sub-fast-magnetosonic regime \citep{2003MNRAS.339..765L}.   

As to the magnetic dissipation in the striped wind zone, its rate can be estimated as follows \citep{2012SSRv..173..341A}.  
In the frame of Earth, the typical thickness of stripes is $l_s=cP/2$, where $P$ is the pulsar period. In  the wind frame, it is $l'_s=\gamma_w cP/2$ and hence the shortest time of stripe dissipation is $t'_s=\gamma_w P/2$. In the Earth frame, the time dilation 
yields $t_s=\gamma_w^2 P/2$.  Once the dissipation is completed $\gamma_{w}=\gamma_{w}^{max}$, which yields the distance to this point as   

\be
     r_{d}= \frac{1}{2} \sigma_M^2 Pc \approx 1.3\times10^7 \lambda_4^{-2}\, \mbox{light days}\,.
     \label{rds}
\ee       
Unless $\lambda_4>300$ this is above the radius of the wind termination shock $r_{ts}\approx 170$ light days ( if we identify it with the inner X-ray ring of the Crab nebula ) and the stripes survive all the way up to the termination shock, where $\gamma_w \ll \sigma_M$.    However, \cite{2003MNRAS.345..153L} has shown that in this case the stripes will dissipate at the termination shock and the post-shock state will be the same as in the case where the stripe dissipation is completed in the wind \citep[see also][]{sironi_spitkovsky_12}. In both these cases, the magnetisation of the post-shock state is determined by the stripe-averaged magnetic field, which vanishes only in the equatorial plane but is the same as the wind field at the boundary with the polar zone. In the transition between the polar and striped zones the magnetization parameter drops rapidly from the high value of the polar zone (see Eq.\ref{sigmaw}) to $\sigma\approx 1$ and then keeps decreasing with the polar angle, but at a lower rate \citep{2013MNRAS.428.2459K}.

%ssssssssssssssssssssssss
\subsection{The nebula}
\label{nebula}

The nebula can be described as a prolate spheroid with the minor axis $a\approx9.5\,\mbox{ly}$ and the major axis $b\approx14\,\mbox{ly}$ \citep{2008ARA&A..46..127H}.  The most spectacular structure of the nebula is its network of thermal filaments.  These are associated with the stellar material ejected during the supernova explosion which left behind the Crab pulsar.  The nonthermal emission is more diffusive but it also exhibits  many features on different length scales. In particular, the radio maps of the nebula show a filamentary structure which is almost identical to network of optical thermal filaments. 

The X-ray images of the inner nebula reveal an equatorial torus-like structure of the radius about 500 light days and a polar jet of a similar size. Inside the torus, there is an oval structure which looks like a beaded neckless around the pulsar, which is called the ``X-ray inner ring''.   Its radius is $r_{ir}\approx 170$ light days.  The ring is often interpreted as the termination shock of the pulsar wind.  Since the pulsar wind is anisotropic, its termination shock is not spherical and it is most extended in the equatorial plane, where it become transverse and forms the so-called Mach belt \citep{KomissarovLyubarsky}. On purely geometrical grounds, one is tempted to identify the ring with the Mach belt, but the nature of its  emission is not clear. \cite{2002ApJ...577L..49H} state that the beards (knots) of the ring ``form, brighten, fade, dissipate, move about, and occasionally undergo outbursts, giving rise to expanding clouds of nebulosity'', though no quantitive parameters of this variability are given. Just the naked eye inspection of the {\it Chandra} images indicates that their size is about one arcsec, which is about the angular resolution of {\it Chandra}, corresponding to the linear scale $\approx 10\,\mbox{ld}$.  One cannot exclude that some knots of the ring are even smaller.            
The ring has no clear identification with features observed in the optical band with HST.  However, the HST images reveal a bright compact knot  at about 0.65 arcsec from the pulsar. This knot is resolved and has the size similar to that of the X-ray knots. The knot is not seen in the radio and X-ray bands \citep{2015ApJ...811...24R}.  In addition to these features, there are also fine bright arc-like filaments (``wisps'') originating in the vicinity of the inner ring and moving away from it with speeds up to $0.7\,c.$. These are seen in the radio, optical and X-ray bands, though the wisps seen in different bands are sometimes displaced relative to each other or do not even have a counterpart in the other band \citep{2004ApJ...615..794B,2008ARA&A..46..127H}.

The observations of both the synchrotron and inverse-Compton emission of the same population of relativistic leptons allows a more-or-less reliable measurement of the nebula energetics. Using a one-zone model of the nebula, \cite{2010A&A...523A...2M} arrive to the mean magnetic field strength of $B\approx0.12\,\mbox{mG}$, with the corresponding magnetic energy density $e_m\approx6\times10^{-10}\,\mbox{erg}\,\mbox{cm}^{-3}$. The lepton population is split into the low-energy sub-population responsible for the radio synchrotron emission and the high-energy population responsible for the optical through the X-ray to the gamma-ray synchrotron emission.  The radio leptons have a hard spectrum, 
$$
dn/d\gamma \propto \gamma^{-p_r},\quad \quad p_r=1.6,\quad  20 < \gamma < 2\times 10^5 
$$ 
and the optical-X-ray leptons have a soft spectrum,   
$$
dn/d\gamma \propto \gamma^{-p_o},\quad p_o=3.2,\quad  4.4\times10^5 <\gamma<3\times10^8 \,.
$$
The total energies of the low and high energy populations are comparable, $E_{l} \approx 3.1\times10^{48}\,$erg and  $E_{h} \approx 2.3\times10^{48}\,$erg respectively. Given the total volume of the nebula, $V\approx 5.7\times10^{56}\,\mbox{cm}^3$, the corresponding total energy density of leptons is $e_p\approx9.4\times10^{-9}\,\mbox{erg}\,\mbox{cm}^{-3}$ and the total pressure (leptons+magnetic field) is $p_{tot}\approx3.6\times10^{-9}$. This total pressure can be balanced by the pressure of  magnetic field  $B=0.3\,\mbox{mG}$, which is close to the standard equipartition value obtained for the Crab nebula \citep{1998ApJ...503..744H}.  

The synchrotron cooling of relativistic electrons leads to a break in the particle spectrum at 
\be 
    \gamma_{b} = (c_2 m_e c^2 B_\perp^2 t_n)^{-1} = 1.6\times10^6 B_{-4}^{-2} t_{n,3}^{-1} \,,
\ee
where $t_{n,3}$ is the nebula age in the units of $10^3\,$years and $c_2=2e^4/3m_e^4c^7 =0.00237$ in CGS units \citep{1970ranp.book.....P}. The corresponding energy of synchrotron photons is 
\be
      \ce_\nu = c_1 B_\perp \ce^2 = 4.4 B_{-4}^{-3} t_{n,3}^{-2}\,\mbox{eV}\,, 
\ee
where $c_1 = 3eh/4\pi m_e^3c^5 = 4.14\times10^{-8}$ CGS units \citep{1970ranp.book.....P}. Thus the spectrum of the optical-X-ray leptons is subject to steepening with the spectral index increasing by unity compared to that of the injection spectrum \citep{1962SvA.....6..317K}, which has to be $a_i = 2.2$.      
 
Since the nebula has a rich structure on different scales, it is natural to expect strong variations of these parameters from feature to feature. Such variations are observed in the MHD simulations of PWN \citep{2013MNRAS.431L..48P,2016JPlPh..82f6301O}, 
with the total pressure exceeding the mean value by up to 5 times at the base of its polar jet and the magnetic field near the termination shock exceeding its mean value by an order of magnitude. Based on these results one may expect the magnetic field in the very inner parts of the Crab nebula to be as high as one mG. 

While it is widely accepted that the high-energy electrons are constantly supplied into the nebula by the pulsar wind,  the nature of its radio electrons is still debated. The observed continuity of the synchrotron spectrum at the transition between the populations has been used to argue that the radio electrons are also supplied by the wind \citep{2011MNRAS.410..381B,2012SSRv..173..341A}. However, this requires a much more prolific particle-production process in pulsar magnetospheres than the one currently accepted.  
Given the results by \cite{2010A&A...523A...2M}, the high and low energy populations contain $N_h\approx 3.8\times 10^{48}$ and  $N_l\approx 3.4\times 10^{51}$ leptons respectively. The corresponding number densities are $n_h\approx6.7\times10^{-9}\,\mbox{cm}^{-3}$ and $n_l\approx6\times10^{-6}\,\mbox{cm}^{-3}$. Given the known age of the nebula $t_n=963\,$yr, the corresponding mean injection rates are $\dot{N}_h \approx1.2\times10^{38}\,\mbox{s}^{-1}$ and  $\dot{N}_l \approx 1.1\times10^{41}\,\mbox{s}^{-1}$ respectively.  Comparing these data with Eq. (\ref{Ndot-theory}) one finds that if only the high energy population leptons are injected by the wind than the required multiplicity parameter $\lambda \approx 10^4$ which is consistent with the theory of pair production in pulsar magnetospheres, whereas if the radio leptons are injected as well then $\lambda\approx 10^7$ \citep{1970ApJ...159L..77S}, which requires a grand revision of this theory   \citep{2012SSRv..173..341A}. Using Eq.\ref{wind-n} for the mean number density of particles in the wind one can check the consistency of these estimates of $\lambda$ with the size of the termination shock. Indeed, at the termination shock the number density of wind particles as measured in the laboratory (Earth's) frame does not change significantly and should match the observed densities in the nebula. Matching it with $n_l$ gives the distance 
$$
    r \approx 70 \, \lambda_7^{1/2}\,\mbox{ld}\,,
$$      
whereas matching with $n_h$ gives 
$$
    r \approx 70\, \lambda_4^{1/2} \,\mbox{ld}\,.
$$
These results are consistent with the observations, which give the equatorial radius of the termination shock $r_{ts}\approx 170\,$ld.  

Using $\lambda=10^7$, we find that upstream of the polar zone of the termination shock  $\gamma_{u}\approx 20$ and $\sigma_{u}\approx 400$. This value of $\gamma_{u}$ is approximately the same as the minimum value $\gamma_{r,min}$ of Crab's radio leptons \citep{2010A&A...523A...2M}, which provides some supports the hypothesis that the radio electrons are supplied by the wind. In the absence of stripes one can rule out the shock acceleration mechanism and hence some other acceleration process has to operate downstream of the shock. According to the results derived in \cite{ 2011MNRAS.414.2017K}, downstream of a highly-magnetised shock  
\be
\gamma_d\approx \sqrt{\sigma_u}/\sin\delta \mtext{and} 
\sigma_d\approx\sqrt{\sigma_u}\gamma_u \sin\delta\,,
\label{downdstr1}
\ee
where $\delta$ is the angle between the upstream velocity and the shock plane.  If the effect of magnetosonic waves is ignored then the upstream parameters can be estimated using Eq.(\ref{sigmaw}), which yields  $ \sigma_u \approx 400 M^{-2/3} \mtext{and} \gamma_u \approx 20 M^{2/3}$ and hence 
$$ 
\gamma_d=20 M^{-1/3} /\sin\delta \mtext{and} \sigma_d=400 M^{1/3} \sin\delta \,.
$$   
We note that $\gamma_d$ and $\sigma_d$ depend rather weakly on the wind Mach number, the parameter which we do not know well but do not expect it to exceed $M\approx10$ by much.  The low wind Lorentz factor obtained in this model is consistent with the radio electrons, which also have low Lorentz factors,  being supplied by the wind.  The high upstream magnetisation rules out particle acceleration at the termination shock and shifts the focus to the processes occurring downstream of the shock instead. The hard spectrum of radio electrons hints the magnetic reconnection mechanism instead but for $\sigma_d>100$ the mechanism yields spectra \citep{2014ApJ...783L..21S, 2015ApJ...806..167G} which are even harder, with $p\approx 1$, than that of the radio electrons\footnote{  Since the synchrotron cooling time of the radio leptons is longer than the age of the Crab nebula their spectrum slope is not affected by the synchrotron energy losses, in contrast to the high energy population.}.  As we have discussed earlier, for this value of the multiplicity parameter the fast-magnetosonic waves produced by the pulsar rotation are expected to dissipate upstream of the termination shock \citep{2003MNRAS.339..765L} and hence to reduce $\sigma_u$ and hence $\sigma_d$, provided they carry a fraction of the total energy flux that exceeds $1/\sigma_u$.

For $\lambda=10^7$ the stripes of the pulsar wind get erased before the termination shock, see Eq. (\ref{rds}),  and hence in the equatorial zone the wind entering the shock should have a unidirectional magnetic field, the magnetisation $\sigma<1$ \citep{2013MNRAS.428.2459K} and the Lorentz factor $\gamma_w=\sigma_M \approx 8\times10^3$.  Because these parameters are very different from those of the polar zone one would expect a different particle acceleration regime (or mechanism) as well, leading to different spectral properties of the accelerated particles.  This hints a possibility of explaining the existence of two different nonthermal populations of the Crab nebula -- the radio electrons are suplied via the polar zone whereas the high-energy electrons via the equatorial striped zone \citep[cf.][]{2013MNRAS.431L..48P,2015MNRAS.449.3149O}.   However the details are not clear.  For example, according to the PIC simulations of relativistic shocks in pair plasma, the magnetisation has to be as low as $\sigma=10^{-3}$ for the traditional Fermi mechanism to operate. Such a low magnetisation is expected only in a very narrow equatorial part of the equatorial zone, with the half-opening angle about one degree for the magnetic inclination angle $\alpha=45^o$ \citep{2013MNRAS.428.2459K}. For the rest of it, the shock is expected only to thermalise the wind particles leading to a Maxwellian-like spectrum with $\gamma_t \approx\gamma_w\approx 8\times10^3$. This would imply a strong emission at the wavelength of  $1-10\,$mm associated with termination shock, which is not seen \citep{2012ApJ...760...96G}.  Finally, the very idea that the two nonthermal populations could be  supplied via different sections of the pulsar wind with very different physical parameters, and presumably shaped via different acceleration mechanisms reintroduces the difficulty of explaining the continuity of the spectrum of the nebula emission.  Moreover, if the radio electrons are supplied via the polar zone only then it should carry about a thousand times more particles than the striped wind zone, which is very hard to explain.

For the more traditional multiplicity $\lambda=10^4$  the radio electrons cannot be supplied by the wind and must have a different origin. For example, the observed filamentary structure of the synchrotron radio emission suggests that they may be closely related to the supernova ejecta shocked by the pulsar wind \citep{2013MNRAS.428.2459K}. However the hard spectral index of the radio emission disfavours the diffusive shock acceleration mechanism.  
As to the termination shock,  Eq.(\ref{sigmaw}) shows that in the polar zone the wind parameters upstream of the shock are $\gamma_u\approx 10^3 M^{2/3}$ and $\sigma_u \approx 10^4 M^{-2/3}$ whereas Eq.(\ref{downdstr1}) shows that downstream of it
$$
\gamma_d \approx 10^3 M^{-1/3}/\sin\delta \mtext{and} \sigma_d \approx 10^4 M^{1/3} \sin\delta\,.
$$ 
These parameters are even more extreme that for $\lambda=10^7$. Moreover, in this case the fast-magnetosonic waves cannot dissipate upstream of the termination shock and hence cannot significantly reduce the wind's $\sigma$. 
 
As to the striped wind zone, Eq.(\ref{rds}) tells us that in this case the stripes survive all the way to the termination shock. As they dissipate at the shock, the spectrum of energised particles depend on the value of parameter 
\be 
    \zeta = 4\pi \lambda \frac{r_{lc}}{r_{ts}}
\ee
\citep{2011ApJ...741...39S}.  For  $\zeta<10$ it is a relativistic Maxwellian-like spectrum whereas a broad power law spectrum is produced only for $\zeta>100$. For the Crab pulsar $r_{lc}\approx 6\times 10^{-8}\,$ld and $r_{ts}\approx 170\,$ld so 
$$
    \zeta \approx 4\times 10^{-4} \lambda_{4} \, .
$$
Thus we expect a  Maxwellian-like spectrum peaking at $\gamma_p \approx \sigma_M \approx 8\times 10^6 \lambda_4^{-1}$. 
The corresponding synchrotron frequency 
$$
         \nu = 5\times 10^{16} B_{-3} \lambda_4^{-2}\, \mbox{Hz} \,.
$$
The inner knot of the Crab nebula is currently associated with the emission of shocked striped wind \citep{2016MNRAS.456..286L}. The knot is seen only in the optical and near IR bands.  Given that the theory predicts  Doppler shift of the knot emission by a factor of few towards higher energies, this requires $\lambda\approx10^5$, which seems quite realistic. More observational data on the knot emission is required to clarify the nature of its emission.    

The above analysis shows a glaring conflict between the observations and the current understanding of relativistic magnetised shocks as particle accelerators. Either the current shock theory based on PIC simulations is completely wrong or the observed spectrum is shaped by other processes. These could be a second-order Fermi mechanism of particle acceleration by plasma turbulence or magnetic reconnection, both occurring inside the nebula.

\section{Papers I \& II: particle acceleration in explosive reconnection events}

In Papers I \& II   we have discussed intensively the properties of particle acceleration during the $X$-point collapse (Paper I),  during the development of the 2D ABC instability and during the merger of  flux tube with zero total current (Paper II). Here we briefly summarize the main results. Overall, the papers investigate particle acceleration during flux merger events in relativistic highly magnetized plasma. 

In Paper I  we studied the initial stages of the current sheet formation - the $X$-point collapse, generalize a classic plasma physics work by \cite{1981ARA&A..19..163S} to the  highly magnetized regime.
Starting from slightly unbalanced X-point configuration, large scale magnetic stresses lead to explosive formation of a current sheet.  During the X-point collapse particles are accelerated by charge-starved electric fields, which can reach (and even exceed) values of the local magnetic field. The explosive stage of reconnection produces non-thermal power-law tails with slopes that depend on the average magnetization $\sigma$.The X-point collapse stage is followed by magnetic island merger that dissipates a large fraction of the initial magnetic energy in a regime of forced magnetic reconnection, further accelerating the particles, but proceeds at a slower reconnection rate.

In Paper II we addressed evolution on somewhat larger scales, conducting a number of numerical simulations  of merging flux tubes with zero poloidal current.
The key difference between this case and the X-point collapse  is that two zero-current flux tubes immersed either in external \Bf\ or external plasma represent a stable configuration - two  barely touching flux tubes, basically, do not evolve -   there is no large scale stresses that force the islands to merge.  When the two flux tubes are pushed together, the initial evolution depends on the transient character of the initial conditions. 

In all the different configurations that we investigated the evolution proceeded according to the similar scenario: initially, perturbation lead to the reconnection of outer field lines and  the formation of common envelope. 
This   initial stage of the merger proceeds very slowly, driven by resistive effects.
With time the envelope grew in size and  a common magnetic envelope develops around the cores. 
 The dynamics changes when the two cores of the flux tube (which carry parallel currents) come into contact. Starting this moment the evolution of the two merging cores resembles the X-point collapse: the cores start to merge explosively and, similarly, later balanced by the forming current sheet. Particle acceleration proceeds here in a qualitatively similar way as in the case of pure $X$-point collapse.
 
The different set-ups allow us to concentrate on somewhat different aspects of particle acceleration. In all the cases that we investigated the efficient particle acceleration always occurs in the region with $ E\geq B$  - by the charge-starved \Efs.  This stage is best probed with the $X$-point collapse simulations, Paper I. In the case of ABC structures and flux tube mergers the
most efficiently particles are accelerated during the initial  explosive stage; during that stage not much of magnetic energy is dissipated. In case of 2D ABC system, this initial stage of  rapid acceleration is followed by a 
 forced reconnection stage; at this stage particles are further accelerated to higher energies, but the rate of acceleration is low. In case of the colliding/merging zero-current flux tubes, the fast dynamic stage is {\it preceded} by the slow resistive stage, when the outer field lines form an overlaying shroud that  pushes the parallel current-carrying cores together. When these parallel current-carrying cores come in contact the evolution proceeds similarly to the unstable ABC case. We stress that {\it the fastest particle acceleration occurs in the beginning of the dynamical stage of the merger} (right away in the X-point collapse simulations, in the initial stage of the instability of the ABC configuration, after the slow resistive evolution in case of colliding/merging flux tubes).

%sssssssssssssssssssssssssssssssssssssssssssss
\section{Gamma-ray flares as reconnection events}
\label{radiation}
%sssssssssssssssssssssssssssssssssssssssssssss

%sssssssssssssssssssss
\subsection{Peak energy}

The fact that the peak energy of the Crab flares is located about the radiation-reaction limit for the synchrotron mechanism suggests that the observed radiation is produced in this regime, whereby the energy gains due to the acceleration mechanism are balanced by the energy losses via the synchrotron radiation.  Accelerated particles are confined to the current sheet  where the magnetic field is weaker compared to the reconnecting magnetic field $B$ outside. Moreover only the component of magnetic field perpendicular to the particle momentum contribute to the synchrotron energy loss. Denoting this component as $B_{a,\perp}=\kappa B$, where $0<\kappa<1$, we can write the energy loss rate by a particle of energy $\ce$ as 
\be
  \frac{d\ce}{dt} = - c_2 B_{a,\perp}^2 \ce^2 \,
\ee    
where $c_2=2e^4/3m_e^4c^7 =0.00237$ in CGS units. The energy gain is via the electrostatic acceleration by the reconnection electric field 
\be
\oder{\ce}{t} = eE_a c \,,
\ee
where $E_a=\eta B$ is the electric field and $0<\eta<1$ is the reconnection rate.  The balance between the two yields the electron energy 
\be
  \ce_{max}=\frac{1}{\kappa} \fracp{\eta ec}{c_2 B} = 1.53 \frac{\eta^{1/2}}{\kappa} B_{-3}^{-1/2}\, \mbox{PeV} \,,
\ee
where $B_{-3}$ is the magnetic field in mG. The corresponding Lorentz factor of flare-producing leptons is 
\be
     \gamma = 3\times10^9 \frac{\eta^{1/2}}{\kappa} B_{-3}^{-1/2} \,.
     \label{b1}
\ee
This is approximately 30 times smaller than the maximal possible  \Lf\ based on the electric potential drop across the pulsar wind given by Eq. (\ref{gammamax1}),  thus supporting the possibility of electrostatic acceleration. Using Eqs.(\ref{pot-drop}-\ref{var-n}) one can estimate the minimum size of the polar wind zone delivering the required potential drop. For 
$\gamma = 3\times10^9$ this is 
\be
  \theta_p^{min} = \left\{\begin{array}{lll}
     15^o & \mbox{for} & n=1\,;\\
     27^o & \mbox{for} & n=2\,.
\end{array} \right.
\ee
These are reasonably small compared to $\theta_p\approx 45^o$ estimated for the Crab pulsar \citep{2008ApJ...680.1378H,2016MNRAS.456..286L}.

The energy of synchrotron photons emitted by the particle of energy $\ce$ in the magnetic field $B$ is 
\be
    \ce_{\nu} = c_1 B_{\perp} \ce^2 \,,
 \label{e_nu}   
\ee  
where $c_1 = 3eh/4\pi m_e^3c^5 = 4.14\times10^{-8}$ CGS units \citep{1970ranp.book.....P}.  It is observed in the simulations \citep{2013ApJ...770..147C} that the highest energy photons are emitted when the accelerated particle is ejected from the current layer into the reconnecting magnetic field.  With this understanding of $B$ in Eq. (\ref{e_nu}) we find the maximum energy of photons emitted from the reconnection region as 
\be 
  \ce_{\nu, max} = \frac{c_1}{c_2} \frac{\eta}{\kappa^2} ec = 158\fracp{\eta}{\kappa^2} \mbox{MeV} \,.
  \label{e-nu-max}
\ee  
In papers I and II we show that for the reconnection driven by macroscopic stresses in highly magnetized plasma, the reconnection speed approaches the speed of light and hence $\eta\to 1$ (Papers I and II). Thus $\kappa\simeq 0.5$ is sufficient to explain the peak photon energy of Crab flares. 

The acceleration time required to reach $\ce_{max}$ is 
\be
   t_{acc} = \frac{\ce_{max}}{\eta eBc} = 2 \frac{1}{\kappa\eta^{1/2}} B_{-3}^{-3/2} \mbox{days} 
   \label{t_acc}
\ee    
whereas the gyration period of an electron with this energy in the reconnecting magnetic field is  
\be 
    t_g = 2\pi\eta t_{acc} \,. 
\ee 
The corresponding gyration radius is  
\be 
     r_g= \eta c t_{acc} \,. 
\ee

%sssssssssssssssssssss
\subsection{Size and energetics of the flaring site}
\label{sec:size}

The acceleration time  sets the lower limit on the size of the ``accelerator'' (the current layer)
\be
    l_{min} =  c t_{acc}  = 2 \frac{1}{\kappa\eta^{1/2}} B_{-3}^{-3/2} \,\mbox{light days} \,. 
    \label{acc-length}
\ee   
Interestingly,  this differs from the gyration radius of the electrons with the energy $\ce_{max}$ only by the factor $1/\eta$.
Assuming that the reconnection region is not moving with relativistic speed,  the  actual size of the active region should fall within the bounds
$$
  l_{min}< l < ct_f\,,
$$
where $t_f$ is the flare duration.  Because $t_f$ is of the order of few days, the last two equations imply that the reconnecting  magnetic field cannot be significantly below  $1\,$mG.  As we discussed in Sec. \ref{nebula}, $B\approx 1\,$mG is too high for most of the nebula but it is not unreasonable for the very inner part of the Crab Nebula.  In fact, direct application of Eq.~(\ref{B_phi}) shows that the wind magnetic field has such a strength at the distance $r=50\sin^n\theta\,$light days, which is significantly smaller compared to the equatorial radius of the termination shock. For $B=1\,$mG the gyration of electrons with the energy ${\ce_{max}}$ is of the order of the size of the active region. This implies that they cannot be trapped by the magnetic islands (ropes) as their size is much smaller compared to the length of the current sheet.   

The total magnetic energy of the flare region is 
\be
       E_m = 7\times10^{38} B_{-3}^2 l_{ld}^3\, \mbox{erg} \,,
\label{fl-energy}
\ee
where $l_{ld}$ is the size of the region in light days. If the fraction $\epsilon_m$ of this energy is converted into the emission with $\ce_\nu>100\,$MeV then the corresponding mean luminosity of the flare is  
\be
        L= 8\times10^{33} \epsilon_m B_{-3}^2\, l_{ld}^3\, t_{f,d}^{-1} \,\mbox{erg}\,\mbox{s}^{-1} \,,
\ee
where $t_{f,d}$ is the flare duration in days.  Since $l$ cannot exceed $ct_f$, the highest possible luminosity is 
\be 
      L_{max} = 8 \times10^{33} \epsilon_m B_{-3}^2\,  t_{f,d}^2 \,\mbox{erg}\,\mbox{s}^{-1} \,.
      \label{L_max}
\ee 
For the April 2011 flare $t_{f,d}=4$ and hence $L_{max} = 1.5 \times10^{35} \epsilon_m B_{-3}^2  \mbox{erg}\,\mbox{s}^{-1} $ whereas the the observed isotropic luminosity above 100 Mev is  $L_{ob}\approx 2\times10^{36} \mbox{erg}\,\mbox{s}^{-1}$. This suggests that either the magnetic field is few times over 1mG or the flare emission is beamed with the beaming factor $f_b = \Delta\Omega/4\pi  \approx L_{max}/L_{ob} \lesssim 0.08\epsilon_m B_{-3}^{2}$ , where  $\Delta\Omega$ is the beam solid angle. 
The angular distribution of particles accelerated in current layers is not isotropic and the degree of anisotropy increases with the particle energy \citep[][Paper I]{2012ApJ...754L..33C}.  This kinetic beaming of accelerated particles is reflected in the anisotropy of their synchrotron emission with the beaming factor near the peak of the emission spectrum $f_b\approx 0.1$, which is consistent with our estimate for the April 2011 flare.

If the reconnecting region is moving relative to the Earth with a relativistic speed then its observed emission can also be beamed due to the Doppler effect. The half-opening angle of the Doppler beam corresponding to the solid angle $\Delta\Omega/4\pi \approx 0.08$ is $\theta_d\approx 35^o$, which corresponds the bulk speed  $v \approx 0.5c$. Such speeds are typical for the inner Crab nebula.    
Even higher speeds are expected in the polar region at the base of the Crab's jet on the theoretical grounds (see Sec.~\ref{nebula}). 

%sssssssssssssssssssss
\subsection{Spectrum of emitting particles}
\label{sec:spectrum}

One can estimate the number of gamma-ray emitting leptons in the flaring region as $N_\gamma\approx Lt_{cool} /\ce_{max}$. Assuming that the cooling time $t_{cool}=t_{acc}$ and using Eqs. (\ref{t_acc}) and (\ref{L_max}), we find 
\be 
    N_\gamma \approx 5.6\times10^{35} \frac{\epsilon_m}{\eta} B_{-3} t_{f,d}^2 \,.    
\ee
(Comparing this with the pair production rate of the pulsar, Eq. (\ref{Ndot-theory}), one can see that it takes a fraction of a second to generate the particles producing the flares.) The corresponding volume density 
\be
   n_\gamma \approx 3\times 10^{-11} B_{-3} t_{f,d}^{-1}\,\mbox{cm}^{-3}
   \label{n-gamma}
\ee
is significantly smaller than the number densities of both the high-energy and low-energy lepton populations in the nebula. This shows that only a tiny fraction fraction of particles in the flaring region are accelerated to the energies required to produce its gamma-ray emission.    

Most studies of particle acceleration in magnetic reconnection come up with a single power law for the nonthermal particles. 
 Our study also indicates a clear possibility of two populations of accelerated particles: a less-energetic population with softer spectrum which is made out of particles trapped inside the magnetic islands (ropes) and a more energetic population with very hard spectrum made out of particles which are free inside the current sheet (Paper I). Only the latter one would produce the flare emission.  
When the synchrotron radiation losses are taken into account one can also expect a pileup of particles near the radiation reaction limit \citep{2012MNRAS.426.1374C}.  Keeping these caveats in mind we will use here the simple model which involves a single power
power law spectrum truncated at the both ends of the energy domain  
\be 
     \frac{dn}{d\gamma} = A \gamma^{-p}, \quad \gamma_{min}\le \gamma \le \gamma_{max} \,.
\ee 
The corresponding spectrum of the synchrotron photons is also a power law $dn_\nu/d\ce \propto \ce^{-a}$, where 
\be
   p= 1+2(a-1) \,.  
\ee
For the April 2011 flare this gives $p=1.54\pm0.24$. Since $1<p<2$ one can estimate  the total number density as 
$$
n_{tot}=A\gamma_{min}^{1-p}/(p-1)
$$ 
and the density of gamma-ray emitting leptons as 
$$
n_\gamma = A\gamma_{max}^{1-p} \,. 
$$
Combining the two, we find 
\be
     \frac{n_{tot}}{n_\gamma} = \frac{1}{p-1} \fracp{\gamma_{max}}{\gamma_{min}}^{p-1}
     \label{nratio}
\ee 
The corresponding mean Lorentz factor 
\be
    <\gamma>  = \gamma_{max} \frac{p-1}{2-p} \fracp{\gamma_{max}}{\gamma_{min}}^{1-p}\,,  
\ee
has to be equal to $\sigma_M$. From the last two equations we find the simple equation 
\be 
   \gamma_{max} = (2-p) \sigma_M \fracp{n_{tot}}{n_\gamma}
\label{gamma-max} 
\ee 
which allows us to check the self-consistency of the model. Substituting the expressions for $\sigma_M$ from Eq.(\ref{sigma-m}), $n_\gamma$ from Eq.(\ref{n-gamma}), using $t_d=4$ and assuming that the wind supples only the high energy leptons ($n_{tot}\approx n_h$), we obtain 
$$
 \gamma_{max} \approx 3.4\times 10^9 \lambda_4^{-1} B_{-3}^{-1}\,,
$$
which is very close to the estimate (\ref{b1}).  If the wind supplies the radio electrons as well then $n_{tot}\approx n_l$ and 
$$
 \gamma_{max} \approx 3\times 10^9 \lambda_7^{-1} B_{-3}^{-1}\,,
$$
which is an equally good result.  Next we can use Eq.(\ref{nratio}) to find 
$$
    \gamma_{min} = \gamma_{max} \left( \frac{n_\gamma}{n_{tot}} \frac{1}{p-1} \right)^\frac{1}{p-1} \,.
$$  
In this equation  $\gamma_{min}$ is very sensitive to the power index.  In the model where the wind supplies only the high-energy population, we find that for $p=1.54\pm0.24$ it varies in the range $\gamma_{min}\in(3\times10^3,5\times10^6)$ with the median value of $5\times 10^5$, corresponding to the optical electrons. If the wind supplies the radio electrons as well then $\gamma_{min}\in(4\times10^{-7},750)$ with the median value of $1.6$, showing that only $p\le 1.54$ are acceptable. Interestingly, the observed spectral index of the radio electrons $p_r = 1.54\pm 0.08$ \citep{1994AZh....71..110K,1997ApJ...490..291B}  is very close to that of the gamma-ray-flare electrons.

%ssssssssssss
\subsection{Plasma parameters}
\label{sec:plasma-par}

The observed spectra of flares provide some information on the plasma magnetisation of the flaring region.  According to the PIC simulations \citep[][Papers I and II,]{2014ApJ...783L..21S, 2015ApJ...806..167G}, $p\approx 1.5$ requires the plasma magnetisation $\sigma$ to be of the order of few tens, with $\sigma>100$ leading to very hard spectra with $p\to 1$ and $\sigma<10$ to soft spectra $p>2$.  

Another important parameter is the electron skin depth $\delta_s = c/\omega_p$, where $\omega_p^2=4\pi e^2 n/(m_e \gamma_t)$ is the plasma frequency  and $\gamma_t$ is the mean Lorentz factor of particles in the reconnection inflow.  Assuming that the flare occurs in plasma whose bulk motion Lorentz factor is of the order of unity but $\sigma \gg 1$, one can estimate $\gamma_t \approx \sigma_M/\sigma$ and hence 
$$
     \omega_p^2 \approx \frac{4\pi e^2}{m_e} \frac{n}{\sigma_M} \sigma \,,
$$ 
which is sensitive to what we assume about the origin of radio electrons.  If they are not supplied by the wind then $n\approx n_h$ and 
\be
     \delta_s \approx \frac{7 \times10^{-3}}{\sqrt{\lambda_4\sigma}}\,\mbox{ld} 
\ee  
and if they are then $n\approx n_l$ and 
\be
     \delta_s \approx \frac{8 \times10^{-6}}{\sqrt{\lambda_7\sigma}}\,\mbox{ld} \,.
\ee  
Thus, the size of the flaring region $ct_{f} \approx \mbox{few}\times10^3 \delta_s$ and $\mbox{few}\times10^6 \delta_s$ respectively.  The corresponding plasma gyration radius 
\be 
      r_g = \frac{m_e c^2 \gamma_t}{e B} = \delta_s/\sigma 
\ee
is even smaller than $\delta_s$.  Thus, the Crab flares undoubtedly involve macroscopic scales.  

As the wind flow is brought to almost a full stop in the flaring site,  $\gamma_w\approx 1$ and  
$$
    \sigma_M = h\sigma =\frac{B^2}{4\pi m_e c^2 n} \,.  
$$ 
This yields an independent estimate of the reconnecting magnetic field strength, namely   
\be 
    B = (4\pi m_e c^2 n \sigma_M)^{1/2} \approx 0.7\,\mbox{mG}\,.   
\ee 
The result is not sensitive to the assumption on the origin of the radio electrons (If the radio electron are supplied by the wind then we should use $n=n_l$ and $\lambda=10^7$ and if they are not then $n=n_h$ and $\lambda=10^7$, which leaves $n\sigma_M$ unchanged.). This is quite close to the lower limit on the magnetic field strength obtained in Sec.\ref{sec:size}.

%sssssssssss
\subsection{Potential-Luminosity relationship and the \Alfven current}

There is Êan important, yet somewhat subtle,  relationship between the parameters of the flares that serve as a confirmation of the basic principles of our approach, as we discuss next.  In addition to particle acceleration the reconnecting \Ef\ 
 also drives (a charge-separated) current.  That induced current creates its own \Bf; particle gyration in the induced \Bf\ can inhibit the beam propagation. 
 These considerations place an upper limit on the total charger-separated current, known as the \Alfven current.
 The relativistic  \Alfven current \citep{1939PhRv...55..425A,1973PhFl...16.1298L} is
\be
I_A = \gamma {m_e c^3 \over e}.
\label{IA}
\ee

In the present model of Crab flares, the reconnection events result in a population of $\gamma$-ray emitting particles. Those particles create electric current, which in turn creates its own \Bf. For the consistency of the model it is required  that the Larmor radius of the particles in their own \Bf\ is sufficiently large. 
Next, using the total flare luminosity and the estimates of the energy of emitting particles we can verify that the current carried by the flare-emitting particles does not violate the \Alfven limit.

 Qualitatively,  the \EM\ power of relativistic outflows can be estimated as  \citep{BlandfordZnajek,2002luml.conf..381B} 
\be 
L_{EM} \approx V^2 /{\cal{R}}
\ee
where $V$ is a typical values of the electric potential produced by the central engine and  ${\cal{R} }\approx 1/ c$ is  the impedance of free space. 
In case of Crab flares, the available potential,
\be
V \sim \sqrt{L /c} \approx 1.7 \times 10^{15} \, {\rm V},
\ee 
corresponds to $\gamma \approx 3.4 \times 10^9$ - very close to the estimate (\ref{b1}). 

Similarly, the electric current associated with  flares
\be
I \approx \sqrt{L  c} \approx 5 \times 10^{22} \, {\rm cgs\, units} =  1.7 \times 10^{13} \, {\rm Ampers}
\ee
is similar to the  \Alfven current (\ref{IA})
 for the required \Lf\ (\ref{b1}):
\be
I_A =  1.7 \times 10^{23} , {\rm cgs\,  units} =  6\times 10^{13} \, {\rm Ampers}
\ee

This is important: the \Lf\ that we  derived from radiation modeling (from the peak energy and duration of flares) nearly coincides with the expected \Lf\ of particles derived from the assumption that the radiated power is similar to the intrinsic \EM\ power  of flares. In a related ``coincidence'', the current required to produce a given  luminosity nearly coincides with the corresponding \Alfven current, calculated using the \Lf\ of the emitting particles.

These coincidences, in our view, are not accidental: they imply that flares are produced by charge-separated flow. This is indeed expected in the acceleration models based on DC-type acceleration. 
This is also consistent with our PIC simulation that show effects of  Òkinetic beamingÓ - highly anisotropic distribution of the highest energy particles.

%fffffffffffffffffffffffffffffffffffffffffffffffffffffff
\section{The likely site}
\label{site}
 
  \subsection{Magnetization in different parts of the nebula}
  
After reviewing the theory and observations of the Crab nebula and the analysis of its gamma-ray flares as magnetic reconnection events, we are now ready to identify the locations in the nebula where this flaring emission may be coming from.  In fact, it is quite clear that for most of the Crab nebula the physical conditions are not favorable for the flares.  Firstly, the mean magnetic field of the nebula is about an order of magnitude below the minimum strength of one mGauss required for the accelerated particles to reach the radiation reaction limit on the typical time scale of the flares (see Sec.\ref{sec:size}).  Secondly, the mean magnetic energy density is about an order of magnitude below that of the synchrotron electrons, which implies that the mean magnetization parameter can be at most $\sigma\approx 0.1$. For such a low magnetization, the reconnection rate can be at most $\eta\approx 0.1$ . This not only increases the required acceleration time even furthermore, see Eq. (\ref{acc-length}), but also reduces the photon energy in radiation-reaction limit by an order of magnitude (see Eq. \ref{e-nu-max}).  Finally, for $\sigma<1$ the spectrum of accelerated particles becomes too soft, with the power index  $p>4$ \citep{2014ApJ...783L..21S, 2015ApJ...806..167G}, whereas the observations suggest a hard power-law spectrum with $p\approx 1.5$ \citep{2012ApJ...749...26B} or even a mono-energetic spectrum \citep{2013ApJ...765...52S}.  If the flares do occur inside the nebula then these must be rather special cites with much stronger magnetic field and much higher magnetization. In principle, strong shocks may amplify the magnetic field.  The RMHD computer simulations of PWN show that secondary shocks are indeed emitted by the highly variable termination shock of the pulsar wind but they are rather weak \citep{2009MNRAS.400.1241C,2013MNRAS.431L..48P}.  Moreover, RMHD shocks cannot change the plasma magnetization from $\sigma\ll 1$ to $\sigma\gg 1$. 
The simulations also indicate that near the termination shock the physical conditions can be more extreme so we now turn our attention to the very vicinity of the shock.

In the analysis of the termination shock and physical parameters of plasma injected by the pulsar wind into the nebula one should differentiate between its polar section, terminating the unstriped polar part of the wind, and its equatorial section, terminating the striped equatorial part of the wind. As \cite{2016MNRAS.456..286L} discuss, the modeling of the inner knot of the Crab Nebula requires that the equatorial flow is weakly magnetized. 
Thus, the physical conditions downstream of the striped section are still not quite suitable for a flaring site. 
Indeed, whether the radio electrons are supplied by the wind or not, the downstream magnetization is very low, with $\sigma \ll 1$. If the radio electrons are supplied by the wind, then its stripes dissipate and its magnetization drops to very low levels before the termination shock. If the radio electrons are not supplied by the wind then the stripes dissipate at the termination shock. As we have discussed, for the parameters of the Crab wind this dissipation leads only to plasma heating (Sec.\ref{nebula}) and the downstream magnetization is too low to make gamma-ray flares via reconnection of the residual magnetic field. The observations of the inner knot of the Crab nebula, which is best explained as a Doppler-boosted emission of the shocked striped wind \citep{2016MNRAS.456..286L}, agree with this verdict -- its optical and IR emission does not correlate with the gamma-ray flares \citep{2015ApJ...811...24R}.

What remains is the poleward zone of the nebula which is filled with the plasma supplied by the unstriped polar section of the pulsar wind. As we discussed in Sec.\ref{nebula}, here the magnetization can be very high $\sigma\gg 1$ leading to the reconnection rate $\eta\approx 1$ and very hard spectra of accelerated particles.  Moreover, the polar region is over-pressured compared to the rest of the nebula due the hoop stress of the azimuthal magnetic field freshly-supplied by the pulsar wind and hence its total pressure and the magnetic field strength can be much higher. Early axisymmetric numerical solutions, based on models of weakly-magnetized pulsar winds, exhibited such magnetic pinch all the way to the outer edge of the nebula \citep[e.g.][]{2009MNRAS.400.1241C}.  However in the recent 3D RMHD simulations of PWN \citep{2013MNRAS.431L..48P,2016JPlPh..82f6301O} such a strong compression is found only at the base of the polar jet, on the scale not exceeding the equatorial size of the termination shock.  Further out the z-pinch appears to be destroyed by instabilities.  

The total potential drop across the polar zone of Crab's pulsar wind is also sufficiently high to accommodate particle acceleration up to $\gamma\approx 3\times10^9$, required by the observations, unless the opening angle of the polar zone is below $15^o$ for $n=1$ and $27^o$ for $n=2$.

At the polar angle decreases below the value of $\theta_p$ separating the striped and polar wind zones, the termination shock  shrinks dramatically in reaction to the much higher magnetization of the polar zone \citep{2012MNRAS.427.1497L,2016MNRAS.456..286L}.  Using the asymptotic solution given in \citet{2016MNRAS.456..286L}, we find that in the polar zone the shock's spherical radius   
\be 
    R(\theta) \simeq \frac{R_o}{\sqrt{2\sigma_u}}  \left\{ 
    \begin{array}{ll}
    {\theta^2}/{2} & \mtext{for} n=1\\
    {\theta^3}/{3} & \mtext{for} n=2
    \end{array}
    \right.
\ee 
where $R_{o}=(S_p/4\pi c p_p)^{1/2}$ and $p_p$ is the total pressure of the polar region.  The equatorial radius of the termination shock can be estimated as  $R_{ts}=(S_p/4\pi c p_n)^{1/2}$, where $p_n$ mean total pressure of the nebula.  Given $p_p\approx 10 p_n$ and $\sigma_u\approx 4\times10^4M^{-2/3}\lambda_4^{-2/3}$ predicted by the ideal MHD model of the pulsar wing, the maximum distance of the termination shock from the pulsar can be well below one light day. Even if fast-magnetosonic waves are responsible for a non-negligible fraction of the initial energy transport in the polar zone and dissipate before reaching the termination shock, the distance is still rather short. A two percent fraction leads to $\sigma_u=50$ and $R<3\,$ld. So we may be dealing with a rather compact region here.

For $\sigma_u \gg 1$ the shock is weak, the speed of the downstream flow is very close to the speed of light and it is highly magnetized. It has to somehow decelerate further in order to match the slow expansion rate of the nebula. \cite{2012MNRAS.427.1497L} analyzed a steady-state axisymmetric model of this polar flow assuming that it is confined by the mean pressure of the nebula, using the approximation of $\theta_p\ll 1$ and assuming $n=1$.  He concluded that the flow converges on the scale 
\be
R_c = \frac{\pi}{\sqrt{6}}\theta_p^2 R_{ts}
\label{r-converg}
\ee
and argued that at this point it will get destroyed by instabilities. For $\theta_p=30^o$, Eq.(\ref{r-converg}) yields $R_c \approx 0.35 R_{ts}$. Given that the external pressure in the polar region can exceed the mean pressure of the nebula by an order of magnitude, $R_c$ can be reduced by a factor of three, leading to $R_c\approx 0.1 R_{ts} \approx 17\,$ld in the case of Crab.       

\subsection{Current  filamentation} 
Recent 3D RMHD simulations of PWN produced by winds with distinctly different polar and equatorial zones \citep{2014MNRAS.438..278P} offer further insight into the dynamics of the polar region. 
As the mildly relativistic  post-shock flow decelerates, it comes into a global causal contact. Once causally connected, the almost purely toroidal magnetic configuration of the nebula flow is liable to current-driven instabilities \cite[e.g.][]{begelman1998,Mizuno:2011aa}.  
As pointed out by \cite{2012MNRAS.427.1497L}, in a highly magnetized expanding post-shock flow, the flow lines are in fact focussed to the axis via the magnetic hoop-stress which can act as an additional trigger for instability.  Thereby the innermost flow reaches the recollimation point first and is expected to experience violent instability and magnetic dissipation.  
\cite{2012MNRAS.427.1497L} hence argues that the $\gamma$-ray flares originate in the polar region at the base of the observed jet/plume.  
On the one hand, the simulations of \cite{2014MNRAS.438..278P} are in agreement with this dynamical picture and show development of modes with azimuthal number $m\sim few$,  that lead to the formation of current filaments at the jet base.  On the other hand, for numerical stability, \cite{2014MNRAS.438..278P} could only study moderate polar magnetizations of $\sigma\le3$ which disfavors rapid particle acceleration.  Even without this technical limitation, for non-vanishing wind power, the axis itself cannot carry a DC Poynting flux.  Because of this, we are interested in flow lines from intermediate latitudes that acquire high sigma due to flow expansion.  

To visualize the current in the simulations, we now cut the domain by spherical surfaces and investigate current-magnitude and magnetization in the $\phi\theta$-plane.    
In general, the current exhibits a quadrupolar morphology, it is distributed smoothly over the polar regions and returns near the equator.  When the test-surface crosses over the termination shock, we observe an increased return-current due to the shock jump of $B_\phi$.  
Although the field is predominantly toroidal in the high sigma downstream region, perturbations from the unstable jet occasionally lead to current filaments and even reversals just downstream of the termination shock.  

\begin{figure}[htbp]
\begin{center}
\includegraphics[height=4.5cm]{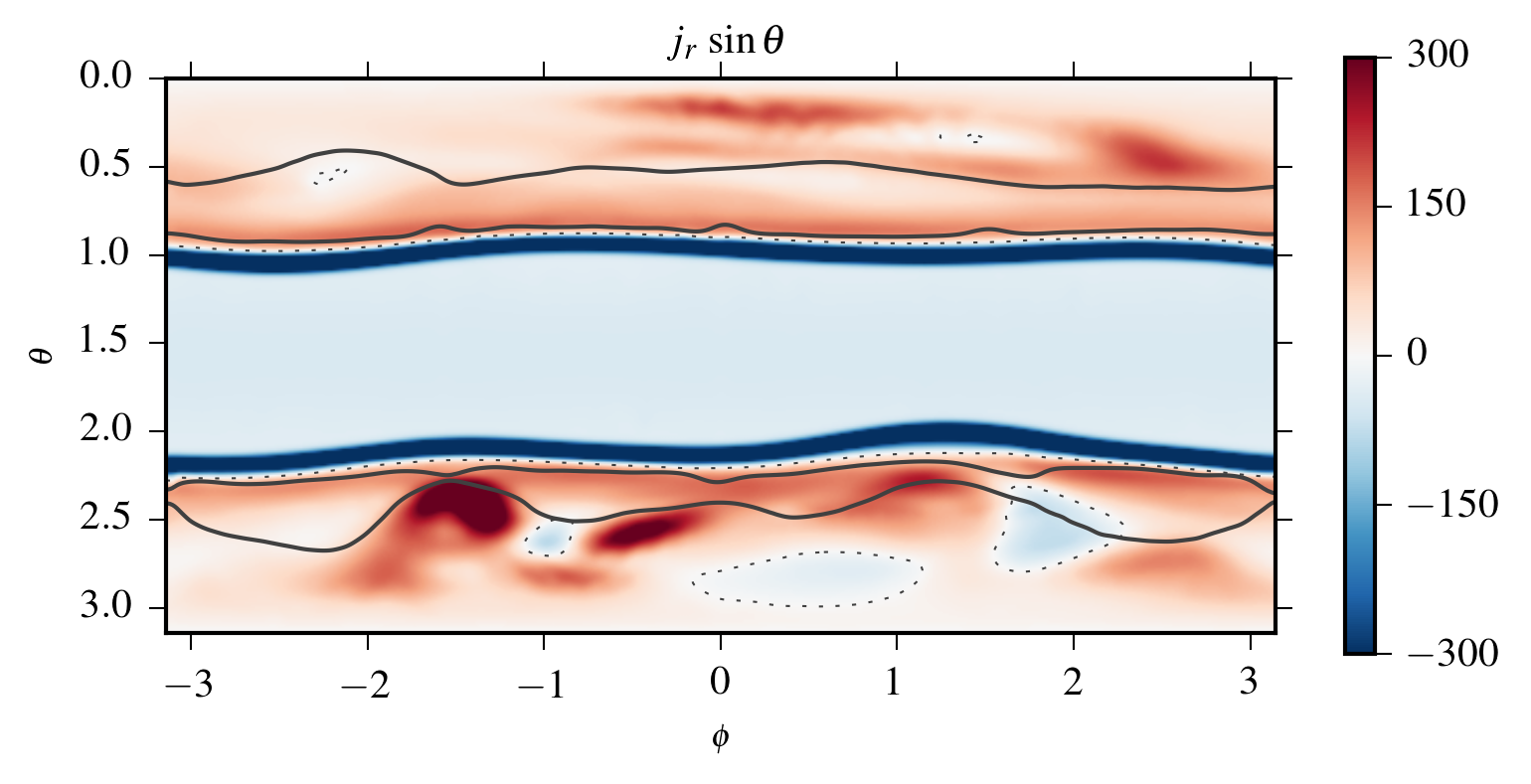}
\includegraphics[height=4.5cm]{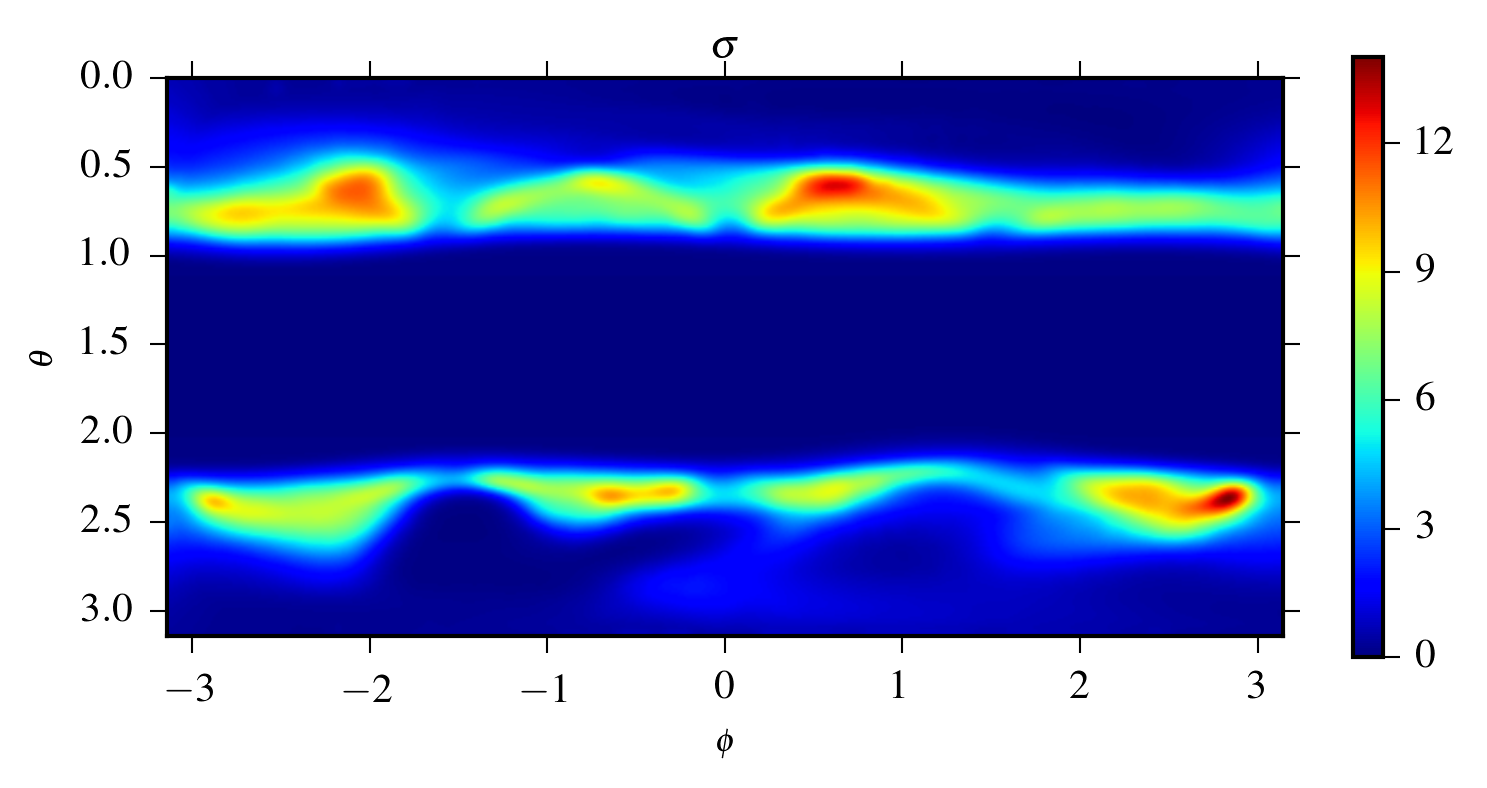}\\
\includegraphics[height=4.5cm]{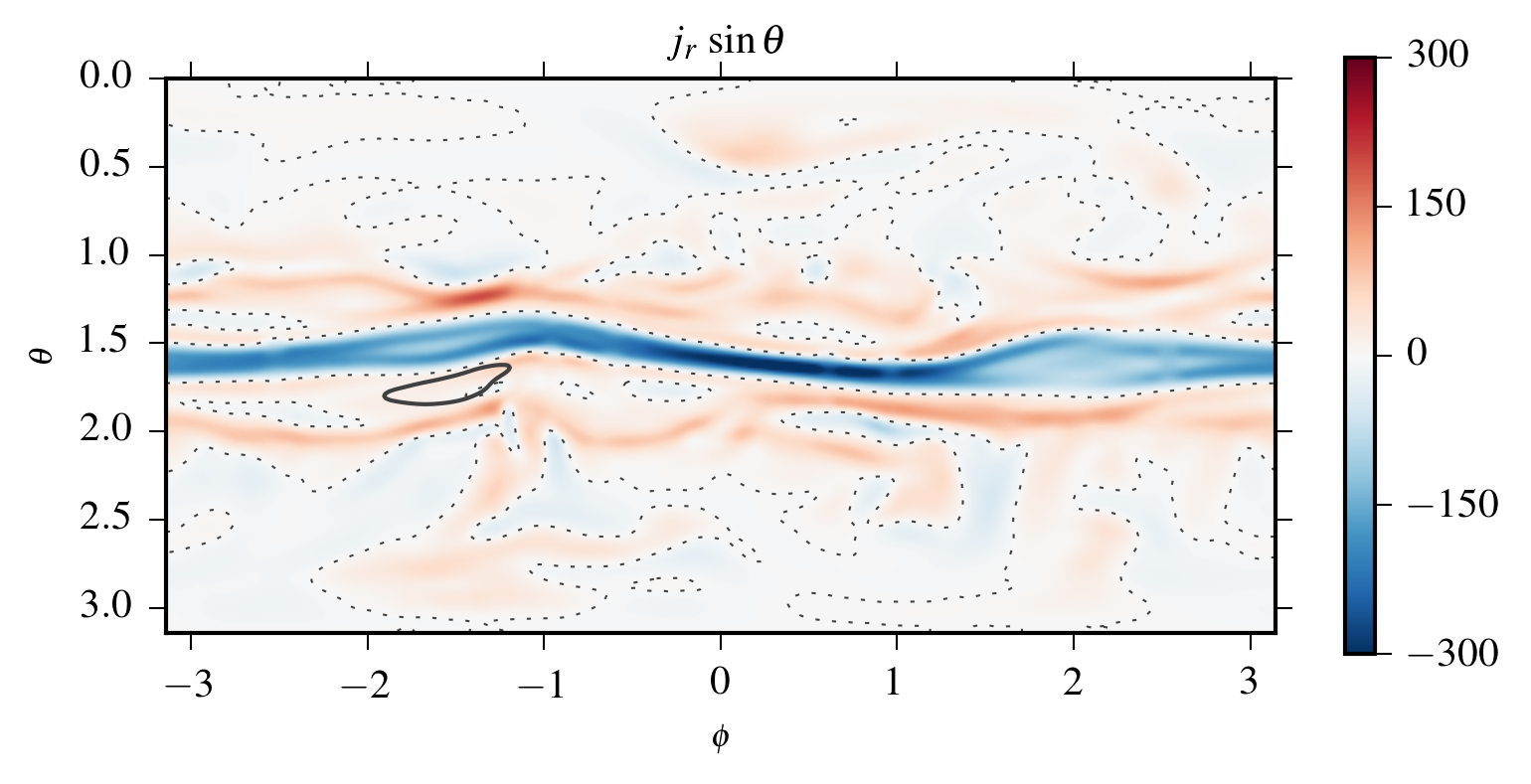}
\includegraphics[height=4.5cm]{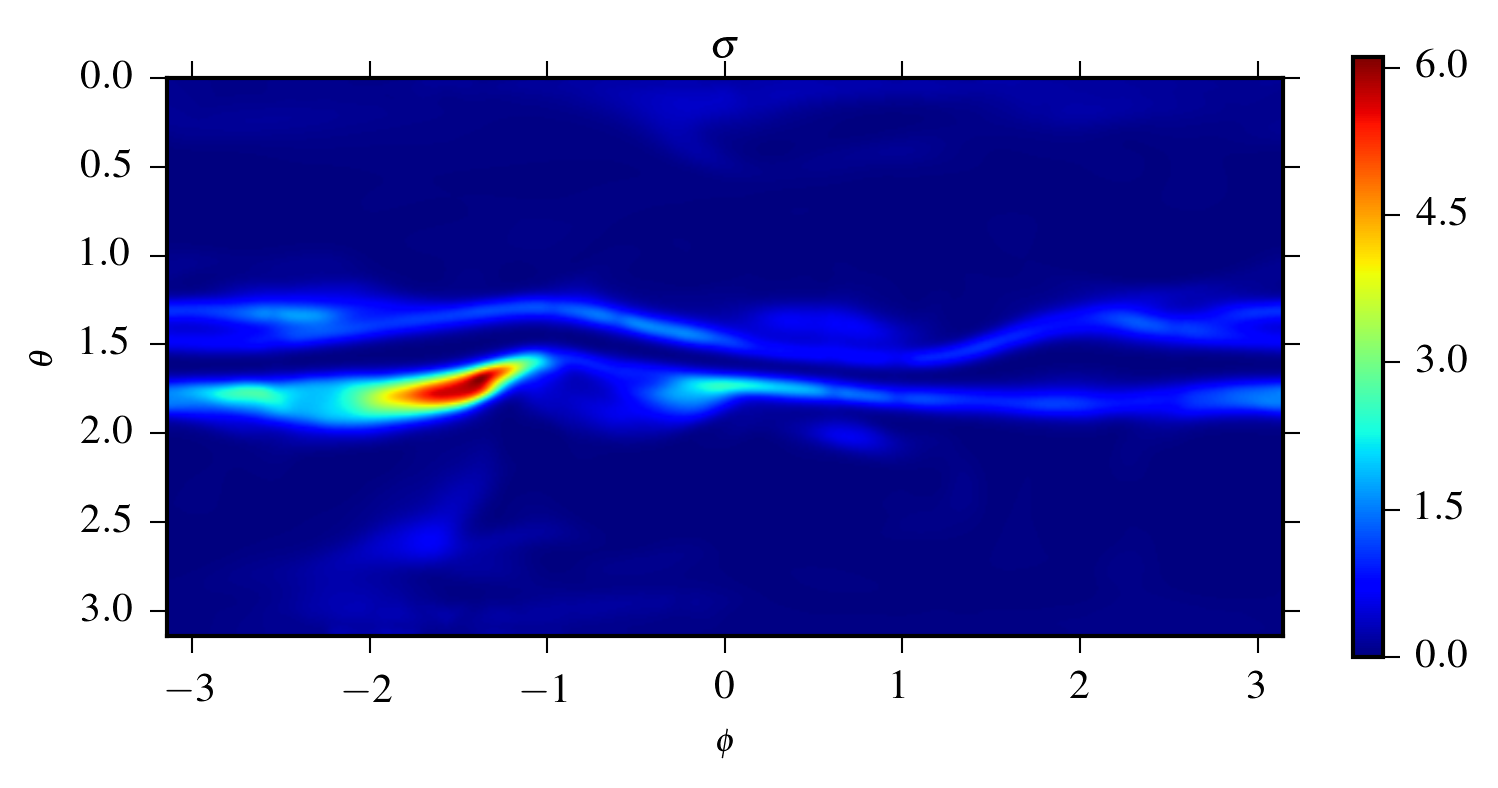}
\caption{Distribution of  the radial current, $j_r\sin\theta$ \textit{(left)} and magnetization $\sigma$ \textit{(right)}.    Dashed contours indicate the sign changes of the current and solid contours indicate levels of $\sigma=4$.  
\textit{Top:} Sphere with $r=10^{17}\rm cm$, 
one can see formation of current filaments (\eg\ regions $\theta \approx 2.5 $ and $\phi \approx -1.5,\, -0.5$). The current direction can even reverse, (the region $\theta \approx 2.7 $,  $\phi \approx -1$). 
\textit{Bottom:} Sphere with $r=3\times 10^{17}\rm cm$.  Here the current is highly filamentary and an equatorial current-sheet has developed.  The average magnetization in the current sheet and in the turbulent flow is $<1$ but magnetically dominated regions can still be obtained ($\sigma\approx6$ at $\theta\approx1.75$, $\phi\approx-1.5$).  
}
\label{fig:jr}
\end{center}
\end{figure}

With increasing distance from the termination-shock, the flow becomes more and more turbulent which leads to increased filamentation and copious sign changes of the radial current.  In the underlying simulations, the magnetization in the turbulent regions has dropped already below unity which disfavors rapid particle acceleration at distances larger than a few termination shock radii.  
It is interesting to point out that although the magnetic field injected in the model decreases gradually over a large striped region of $45^\circ$, an equatorial current sheet develops none the less in the nebula.  This can be seen in the lower panel of  Fig.\ref{fig:jr} which corresponds to a surface outside of the termination shock at $r=3\times 10^{17}\rm cm$.  

\begin{figure}[htbp]
\begin{center}
\includegraphics[width=0.7\textwidth]{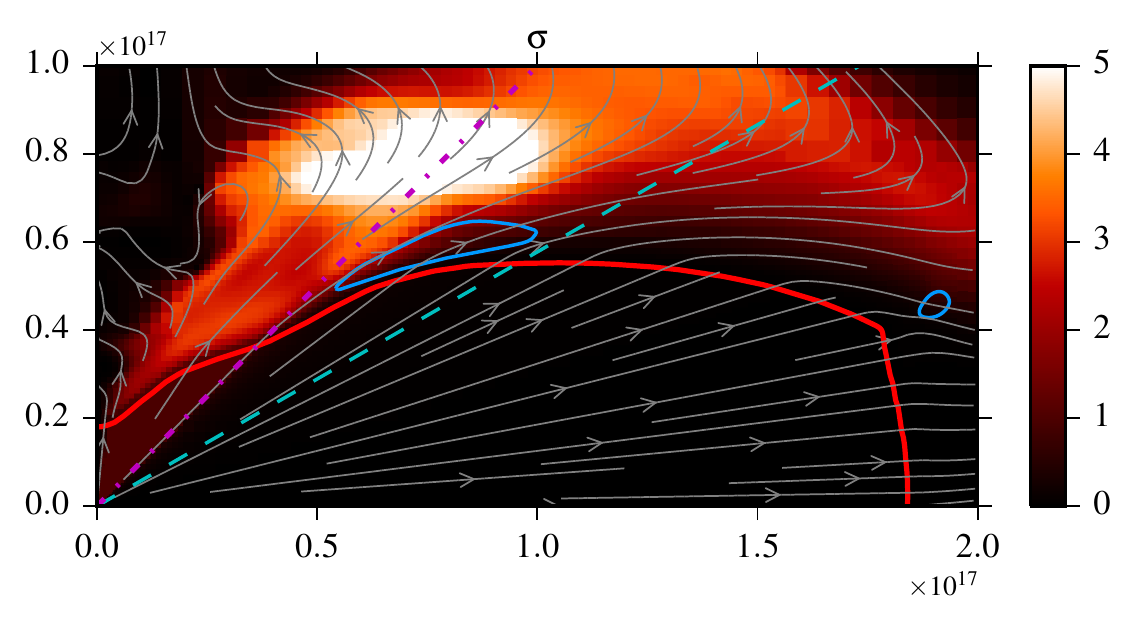}
\caption{Magnetization $\sigma$ and instantaneous streamlines near the termination shock in RMHD simulations of the Crab nebula \citep{2014MNRAS.438..278P}.  The dot-dashed straight line shows the separation of the polar and striped zones of the pulsar wind. The dashed straight line is the line of sight. The solid red line shows the termination shock and the solid blue loop between the dashed and dot-dashed lines shows the region of Doppler-beamed emission associated with the inner knot of the nebula. The polar beam corresponds to the streamlines originating from the inner part of the termination shock located to the left of the intersection with the dot-dashed line.}
\label{fig:rmhd}
\end{center}
\end{figure}

Figure~\ref{fig:rmhd} shows the results for a wind with $\theta_p=45^o$ and $\sigma_u \approx 1$ in the polar zone. As one can see, the polar flow does indeed converge towards the polar axis with $R_c\approx 0.5 R_{ts}$. Moreover, a significant fraction of the shocked equatorial part of the wind also converges, in  agreement with the earlier 2D simulations of weakly magnetized pulsar winds. Furthermore, the polar flow does not look at all as a quasi-stationary axisymmetric jet. It is highly disturbed already on the scales $R \ll R_c$, with even a negative radial velocity in one region.

\begin{figure}[htbp]
\begin{center}
\includegraphics[height=10cm]{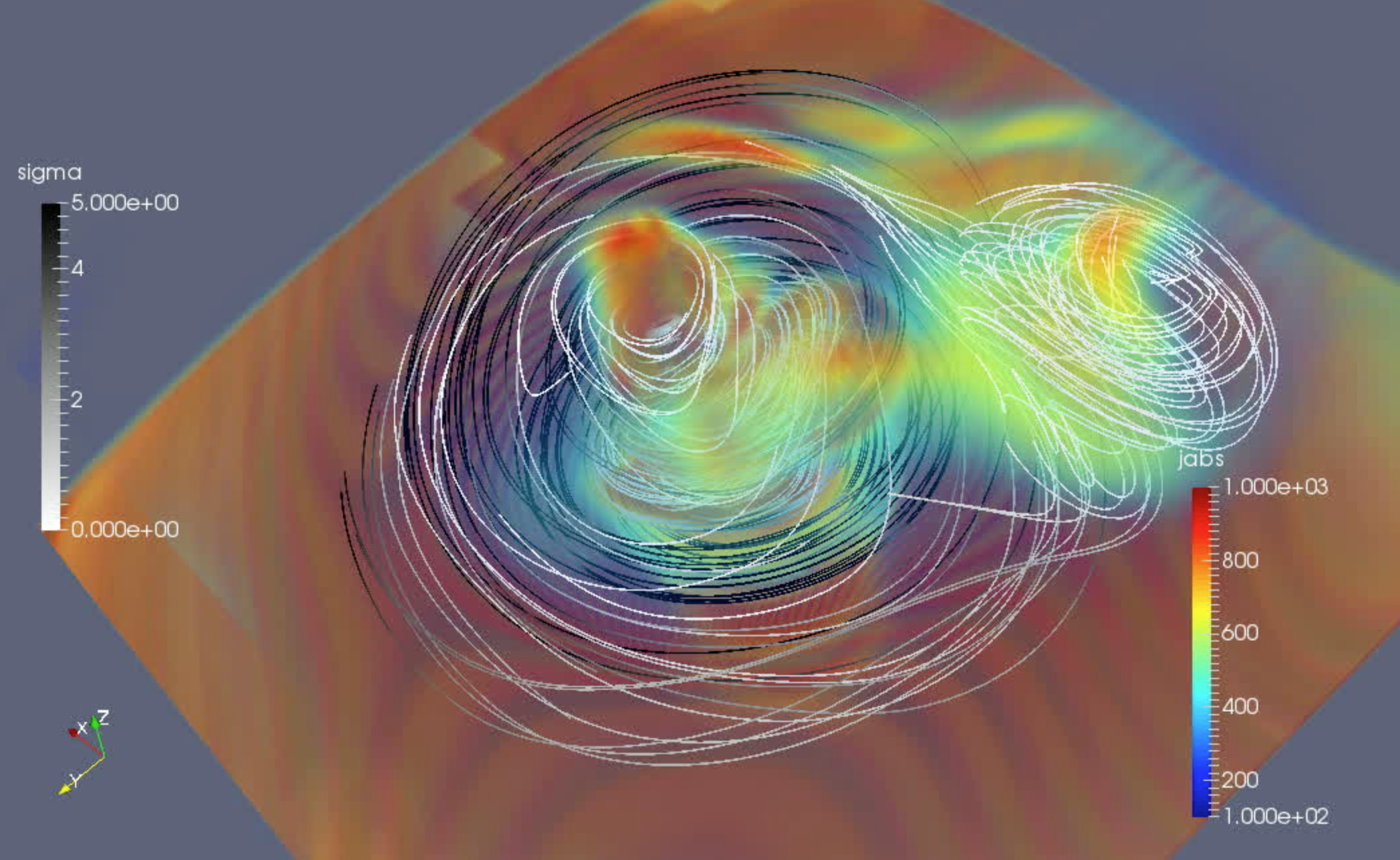}
\caption{3D volume rendering showing current filamentation of the polar beam just downstream of the termination shock.  The shock surface is indicated as the orange plane and we draw field-lines shaded from white $(\sigma=0)$ to black ($\sigma=5$).  
One clearly sees two current filaments producing structures similar to magnetic flux tubes.  As discussed in \cite{2014MNRAS.438..278P}, streamlines from intermediate latitudes reach the axis behind this inner violently unstable region and form a plume-like outflow of moderate velocity $v\approx0.7c$.  
}
\label{fig:filamentation}
\end{center}
\end{figure}

To better understand the geometry of the current and magnetic field, we display a representative volume rendering of the polar region in Fig. \ref{fig:filamentation}, which shows the complicated structure of field lines in the polar zone on the scales of $r\in(0.1,0.3)R_{ts}$. n the rendering  in Fig. \ref{fig:filamentation} one can see the violently unstable polar beam embedded into the more regular high-$\sigma$ region comprised of toroidal field lines.  
The plume forms downstream of this structure and is also strongly perturbed.  A part of the disrupted plume approaches the termination shock as a flux-tube.  
The presence of such a configuration where two flux tubes can come together very close to the high sigma region lets us speculate that Crab flares might originate when the right geometry (e.g. parallel flux tubes) coincides with high magnetization as present in the nebula even for moderate wind magnetization.  
For higher magnetizations of the polar beam, the mechanism described by \cite{2012MNRAS.427.1497L} could directly act also without first having to rely on enhancement of $\sigma$ via flow expansion.  Extrapolating from the moderate sigma simulations where the polar beam is highly unstable and forms a filamented current, this seems feasible at the very least.  

The overall impression is that the converging part of equatorial flow creates a highly dynamic ``wall'' that almost stops and crushes the polar flow, which looses its regular structure and develops numerous current sheets well before the convergence point. This must be followed by a rapid forced magnetic reconnection akin the one observed in the simulations of shocks in plasma with magnetic stripes \citep{2011ApJ...741...39S} and in our simulations of magnetic reconnection driven by macroscopic stresses (papers I and II). The conversion of magnetic energy into the energy of particles during the magnetic reconnection increases the plasma compressibility and this may be the mechanism ultimately responsible for the observed slowing down of the polar beam. 

Summarizing, the poleward regions downstream of the shock terminating the unstriped section of the pulsar wind seems to be the only promising location for the gamma-ray flares of the Crab nebula. 
 This is a rather compact central region and in the maps of the Crab nebula it occupies the area inside the X-ray ring of the Crab nebula. The flare region should be located at (or extend up to) intermediate polar angles in the range of $10-30$ degrees.  The required acceleration potential  within the region is also a  substantial fraction (few \%) of the total potential of the wind. The flare region  has a macroscopic size, orders of magnitude  larger than the corresponding skin depth.

\section{Discussion}

\subsection{Crab flares: What have we learned?}

So far most of the studies of relativistic magnetic reconnection focused on the evolution of a very long \footnote{In numerical simulations, periodic boundary conditions allows to talk about infinitely long current sheets.} current sheet with thickness of the order of the skin depth of plasma in this initial configuration. Such a small thickness  is required to ensure a rapid development of collisionless tearing instability and hence a rapid onset of reconnection.  In Sec.\ref{sec:plasma-par} we have shown that in the case of Crab flares the current sheet must be at least few thousands, and may be even millions, of skin depths long. At first it looks as this is a very special configuration which can be arranged only in the control environment of physical or numerical laboratory.  However, both our fluid and PIC simulations reported in Papers I and II, show that there is no need in arranging such a configuration. The macroscopic stresses in highly magnetised relativistic plasma drive rapid restructuring of current-sheet-free initial configurations and development of thin current sheets on the length scale of the magnetic field variation in these configurations. The thickness of these current sheets is determined by the point at which  the particle acceleration turns on, which is the point of electric charge starvation. Initially the current sheets may be rather short or even non existent, as in the case of the X-point collapse (Paper I), but once formed they grow at about the speed of light. The reconnection rate is also comparable to the speed of light and hence the whole configuration evolves in a highly unsteady, explosive fashion. This is one property which makes the relativistic magnetic reconnection an attractive model for Crab's flares.   

As the energy of gamma-ray photons peaks near, or even above, the radiation-reaction limit (Eq.\ref{e-nu-max}) the emitting electrons must be accelerated by an electric field of the order of the magnetic field all the way until the reach the energy required to produce this emission.  The relativistic magnetic reconnection allows acceleration  (electrostatic) of the sort as the strength of reconnection electric field in the current sheet can be very close to the strength of the reconnecting magnetic field outside.  Moreover, our PIC simulations (Papers I and II) show that while most involved particles are ``unlucky'' as they become trapped inside magnetic ``ropes'' developed via the tearing instability of the current sheet, where their acceleration becomes interrupted, there are ``lucky'' particles which are not affected by the ropes as their gyration radius  radius exceeds the rope radius.  These are continuously accelerated until they fulfil the total electrostatic potential available in the current sheet  or reach the radiation reaction limit.  To become lucky, a particle has to enter the current sheet via microscopic X-points.
 
 Reviewing the theoretical models of pulsar winds and both the theory and the observations of the Crab Nebula we have identified the region in the nebula where the gamma-ray flares are likely to originate. This is a compact polar zone at the base of the Crab jet where the plasma magnetisation can be sufficiently high and the magnetic field sufficiently strong.  Its size is not expected to exceed the equatorial radius of the termination shock and hence in the plane of sky it should be located within the inner X-ray ring of the Nebula.

\subsection{Crab flares: open questions}

Strong gamma-ray flares of the Crab nebula do not follow one straight after another but are separated by approximately one year instead. \citet{2013ApJ...765...52S} identified a similar number of lower amplitude and longer duration events, which they called ``waves'', but even if we add them to the flares proper, we are still dealing with rather rare events. One can also approach the question of their frequency on the basis of involved energetics.   In order to have sufficient potential drop, the polar zone should have $\theta_p>15^o$ for the wind model with $n=1$ and $\theta_p>27^o$ for $n=2$.  Based on the observed properties of the inner knot of the Crab nebula, \citet{2016MNRAS.456..286L} give $\theta_p<45^o$.    We can use these to set the limits of the total energy supplied into the  polar zones during one year:  $2\times10^{43}\mbox{erg}<E_p<7\times10^{44}\mbox{erg}$  for $n=1$ and  $2\times10^{43}\mbox{erg}<E_p<3\times10^{44}\mbox{erg}$ for $n=2$.  These can be compared with the total energy  $E_\gamma\approx 5\times10^{40}$erg radiated during the gamma-ray evens for the same period (estimated using Eq.(\ref{fl-energy}). Thus only a tiny fraction of the magnetic energy dissipated in the polar region results in flares. The kinetic beaming alone cannot explain this disparity as its beaming factor is found to be only about $f_b=0.1$ (see Sec.\ref{sec:size}). However, there are number of other possible reasons.

The potential drop across the whole polar zone is only just above the one required to explain the peak energy of the flares. If typical current sheets are rather small compared to the size of the polar zone, their potential drops may be significantly lower, resulting in a lower maximum energy of accelerated particles and hence a lower peak energy of the flares. Events with $\ce_\nu< 100\,$MeV would be much less prominent due to the much higher mean flux from the nebula at these energies.                  

The typical plasma magnetisation $\sigma$ could be too high, leading to very hard spectra. In order to understand why this could  be a problem, let us consider a mono-energetic spectrum. In this case, all particles would have the Lorentz factor $\gamma\approx\sigma_M=8\times10^6 \lambda_{4}^{-1}$, which is well below $\gamma = \mbox{few}\times10^9$ required for the flare electrons. In order to illustrate this for the case of a power-law spectrum, we can rewrite Eq.(\ref{gamma-max}) as 
\be
    \gamma_{max}^p= \sigma_M \frac{2-p}{p-1} \gamma_{min}^{p-1} 
\ee 
and consider $p=1.1$ as an example of very hard spectrum corresponding to $\sigma>100$. Even for $\lambda_4=1$ and $\gamma_{min}=10^5$ this gives $\gamma_{max}\approx 4\times10^7$ only. In the magnetic field of strength  $B=1\,$mG the synchrotron emission of such particles would not extend much above the cutoff energy of  $\ce_\nu\approx30\,$keV.

In the polar zone, the shocked wind plasma flows predominantly along the polar axis, with the initial half-opening angle $\theta\approx\theta_p$. The polar axis of the Crab pulsar makes the angle of about $60^o$ to the line of sight \citep{2004ApJ...601..479N}. Downstream of the termination shock we expect its speed to be very close to the speed of light, with $\gamma\gg1$, and its emission strongly beamed in the direction of motion. As the result, its observed emission is Doppler-dimmed. Although, the flow eventually slows down to the characteristic speed $v_j\approx 0.4c$ of Crab's polar jet  \citep{2002ApJ...577L..49H}, we may still expect that most of the emission associated with magnetic reconnection in the polar zone is beamed away from Earth.    

Thus, the observed gamma-ray flares seem to be associated with rather exceptional circumstances when the reconnection region is rather large in terms of its potential drop, its magnetisation is rather low, presumably due to some previous phases of reconnection, and its velocity is either strongly reduced or directed towards the line of sight, e.g. due to interaction with its highly-disturbed low-sigma  surrounding.  High-resolution  RMHD or PIC simulations of the polar region are required to clarify this issue. 

While the shock model of \citet{1984ApJ...283..710K} allows a simple prescription for the emission of the Crab Nebula, the magnetic reconnection does not. It remains to be seen what kind of integral particle spectra can be produced in this model. The model will have to incorporate two-component (the polar zone plus the equatorial zone) structure of pulsar winds as they inject plasmas with very different physical parameters. Moreover, due to the anisotropy of the wind power the striped component is expected to carry most energy of the energy and the magnetisation of plasma it injects into the nebula is quite low $\sigma<1$. For such a magnetisation the spectrum of nonthermal electrons accelerated via magnetic reconnection is expected to be very soft $p>4$, in fact too soft to agree with the observations. This suggests that another acceleration mechanism is at work there, presumably the second-order Fermi mechanism involving plasma turbulence (citations).  

Based on the results of PIC simulations of shocks in relativistic magnetised plasma the termination shock of pulsar winds only thermalises plasma (except the very narrow section near the equator where the first-order Fermi mechanism may operate). This conclusion makes  observational identification of the termination shock and studying of its broad band spectrum particularly important.  So far we have two candidates: the X-ray ring of the Crab nebula and its inner knot. In both cases the broadband spectrum is not known. The inner ring may correspond to the narrow equatorial section of the termination shock where the emitting particles are accelerated via the Fermi mechanism, whereas the inner knot corresponds to the high-latitude part of the termination shock, terminating the striped section of the pulsar wind, where the theory predicts a relativistic Maxwellian spectrum of the emitting particles.        

%sssssssssssssssssssss
\subsection{Implications for other high energy sources: acceleration by  shocks and/or magnetic   reconnection}
\label{Implication}

The Crab Nebula is the paragon of  the high energy astrophysical source.  Both the observational and theoretical studies of the Crab Nebula have made a very strong impact on the development of high energy astrophysics in general.  Given the proximity of this object and the unprecedented level of detail it offers, one would expect that it will retain its special place.   Many other astrophysical phenomena involve flows of highly-magnetised relativistic plasma.  Most of all, these are Active Galactic Nuclei and presumably Gamma Ray Bursts as well  \citep[\eg][]{BlandfordZnajek,Usov92,2006NJPh....8..119L,2007MNRAS.382.1029K,2004ApJ...611..380T}, which involve production of relativistic jets from compact magnetised rotators, black holes and neutron stars.  

Most of the emission coming from AGN jets is non-thermal, which makes the issue of non-thermal particle acceleration central in their studies. For a long time it was believed that such particles are accelerated at shocks, which naturally emerge in supersonic flows.   However, the recent PIC simulations of relativistic plasma place severe constrains on the shock-acceleration mechanism in magnetised relativistic plasma -- it operates only in the case of weak magnetisation, $\sigma < 10^{-2}$.  Although these results are often ignored and the shock model is still in use,  the high-energy  astrophysics is in a mini-crisis.   Relativistic magnetic reconnection may be offering a way out this crisis both in the case of AGN  \citep[e.g.][]{2002luml.conf..381B,2003NewAR..47..513L,2011SSRv..160...45U,2015MNRAS.450..183S} and GRBs \citep[e.g.][]{2003ApJ...597..998L,2006NJPh....8..119L}. It is quite possible that the high energy flares of AGNs as well as the spikes in the light curves of GRBs are similar in nature to the flares of Crab Nebula \citep{Lyutikov:2006a,2010MNRAS.402.1649G,2012MNRAS.426.1374C} and hence exploring the magnetic reconnection model for the emission of the Crab Nebula will have important implications in the astrophysics of AGNs and GRBs once again.

%\citep{2011PlPhR..37..118Z}:  both drift modes and tearing-like instabilities can develop. 

%A thin current layer can also bend along the z-direction due to the relativistic drift-kink instability (e.g., Zenitani \& Hoshino 2005; Liu et al. 2011),

\acknowledgements
We would like  to thank Roger Blandford,  Krzystof Nalewajko,  Dmitri Uzdensky and Jonathan Zrake for stimulating discussions.
This research was supported via NASA grant NNX12AF92G, NSF  grant AST-1306672 and DoE grant 108483.
OP is supported by the ERC Synergy Grant ``BlackHoleCam -- Imaging the Event Horizon of Black Holes'' (Grant 610058). SSK is supported by the STFC grant ST/N000676/1. The simulations were performed on XSEDE resources under
contract No. TG-AST120010, and on NASA High-End Computing (HEC) resources through the NASA Advanced Supercomputing (NAS) Division at Ames Research Center. ML would like to thank   for hospitality Osservatorio Astrofisico di Arcetri and Institut de Ciencies de l'Espai, where large parts of this work were conducted. SSK thanks Purdue University for hospitality during sabbatical in 2014 when a significant part of this work was carried out. 
  
 \clearpage
 \bibliography{/Users/maxim/Home/Research/BibTex,./additions} 
\bibliographystyle{jpp}

   \end{document}